\documentclass[ALICE,manyauthors]{cernphprep}
\usepackage[comma,square,numbers,sort&compress]{natbib}
\usepackage{hyperref}
\usepackage{lineno}
\usepackage{xspace}
\usepackage{multirow}
\usepackage{color}
\usepackage[T1]{fontenc}
\usepackage{euscript}

\begin{document}
%

\newcommand{\pp}           {pp\xspace}
\newcommand{\ppbar}        {\mbox{$\mathrm {p\overline{p}}$}\xspace}
\newcommand{\XeXe}         {\mbox{Xe--Xe}\xspace}
\newcommand{\PbPb}         {\mbox{Pb--Pb}\xspace}
\newcommand{\pA}           {\mbox{pA}\xspace}
\newcommand{\pPb}          {\mbox{p--Pb}\xspace}
\newcommand{\AuAu}         {\mbox{Au-two-Au}\xspace}
\newcommand{\dAu}          {\mbox{d--Au}\xspace}

\newcommand{\s}            {\ensuremath{\sqrt{s}}\xspace}
\newcommand{\snn}          {\ensuremath{\sqrt{s_{\mathrm{NN}}}}\xspace}
\newcommand{\pt}           {\ensuremath{p_{\rm T}}\xspace}
\newcommand{\meanpt}       {$\langle p_{\mathrm{T}}\rangle$\xspace}
\newcommand{\ycms}         {\ensuremath{y_{\rm CMS}}\xspace}
\newcommand{\ylab}         {\ensuremath{y_{\rm lab}}\xspace}
\newcommand{\etarange}[1]  {\mbox{$\left | \eta \right |~<~#1$}}
\newcommand{\yrange}[1]    {\mbox{$\left | y \right |~<~#1$}}
\newcommand{\dndy}         {\ensuremath{\mathrm{d}N_\mathrm{ch}/\mathrm{d}y}\xspace}
\newcommand{\dndeta}       {\ensuremath{\mathrm{d}N_\mathrm{ch}/\mathrm{d}\eta}\xspace}
\newcommand{\avdndeta}     {\ensuremath{\langle\dndeta\rangle}\xspace}
\newcommand{\dNdy}         {\ensuremath{\mathrm{d}N_\mathrm{ch}/\mathrm{d}y}\xspace}
\newcommand{\Npart}        {\ensuremath{N_\mathrm{part}}\xspace}
\newcommand{\Ncoll}        {\ensuremath{N_\mathrm{coll}}\xspace}
\newcommand{\dEdx}         {\ensuremath{\textrm{d}E/\textrm{d}x}\xspace}
\newcommand{\RpPb}         {\ensuremath{R_{\rm pPb}}\xspace}

\newcommand{\nineH}        {$\sqrt{s}~=~0.9$~Te\kern-.1emV\xspace}
\newcommand{\seven}        {$\sqrt{s}~=~7$~Te\kern-.1emV\xspace}
\newcommand{\twoH}         {$\sqrt{s}~=~0.2$~Te\kern-.1emV\xspace}
\newcommand{\twosevensix}  {$\sqrt{s}~=~2.76$~Te\kern-.1emV\xspace}
\newcommand{\five}         {$\sqrt{s}~=~5.02$~Te\kern-.1emV\xspace}
\newcommand{\twosevensixnn}{$\sqrt{s_{\mathrm{NN}}}~=~2.76$~Te\kern-.1emV\xspace}
\newcommand{\fivenn}       {$\sqrt{s_{\mathrm{NN}}}~=~5.02$~Te\kern-.1emV\xspace}
\newcommand{\LT}           {L{\'e}vy-Tsallis\xspace}
\newcommand{\GeVc}         {Ge\kern-.1emV/$c$\xspace}
\newcommand{\MeVc}         {Me\kern-.1emV/$c$\xspace}
\newcommand{\TeV}          {Te\kern-.1emV\xspace}
\newcommand{\GeV}          {Ge\kern-.1emV\xspace}
\newcommand{\MeV}          {Me\kern-.1emV\xspace}
\newcommand{\GeVmass}      {Ge\kern-.2emV/$c^2$\xspace}
\newcommand{\MeVmass}      {Me\kern-.2emV/$c^2$\xspace}
\newcommand{\lumi}         {\ensuremath{\mathcal{L}}\xspace}

\newcommand{\ITS}          {\rm{ITS}\xspace}
\newcommand{\TOF}          {\rm{TOF}\xspace}
\newcommand{\ZDC}          {\rm{ZDC}\xspace}
\newcommand{\ZDCs}         {\rm{ZDCs}\xspace}
\newcommand{\ZNA}          {\rm{ZNA}\xspace}
\newcommand{\ZNC}          {\rm{ZNC}\xspace}
\newcommand{\SPD}          {\rm{SPD}\xspace}
\newcommand{\SDD}          {\rm{SDD}\xspace}
\newcommand{\SSD}          {\rm{SSD}\xspace}
\newcommand{\TPC}          {\rm{TPC}\xspace}
\newcommand{\TRD}          {\rm{TRD}\xspace}
\newcommand{\VZERO}        {\rm{V0}\xspace}
\newcommand{\VZEROA}       {\rm{V0A}\xspace}
\newcommand{\VZEROC}       {\rm{V0C}\xspace}
\newcommand{\Vdecay} 	   {\ensuremath{V^{0}}\xspace}

\newcommand{\ee}           {\ensuremath{e^{+}e^{-}}} 
\newcommand{\pip}          {\ensuremath{\pi^{+}}\xspace}
\newcommand{\pim}          {\ensuremath{\pi^{-}}\xspace}
\newcommand{\kap}          {\ensuremath{\rm{K}^{+}}\xspace}
\newcommand{\kam}          {\ensuremath{\rm{K}^{-}}\xspace}
\newcommand{\pbar}         {\ensuremath{\rm\overline{p}}\xspace}
\newcommand{\kzero}        {\ensuremath{{\rm K}^{0}_{\rm{S}}}\xspace}
\newcommand{\lmb}          {\ensuremath{\Lambda}\xspace}
\newcommand{\almb}         {\ensuremath{\overline{\Lambda}}\xspace}
\newcommand{\Om}           {\ensuremath{\Omega^-}\xspace}
\newcommand{\Mo}           {\ensuremath{\overline{\Omega}^+}\xspace}
\newcommand{\X}            {\ensuremath{\Xi^-}\xspace}
\newcommand{\Ix}           {\ensuremath{\overline{\Xi}^+}\xspace}
\newcommand{\Xis}          {\ensuremath{\Xi^{\pm}}\xspace}
\newcommand{\Oms}          {\ensuremath{\Omega^{\pm}}\xspace}
\newcommand{\degree}       {\ensuremath{^{\rm o}}\xspace}

\newcommand{\dd}     {\mathrm{d}}
\newcommand{\dbar}   {$\overline{\mathrm{d}}$}
\newcommand{\vtwo}   {$v_{2}$\xspace}
\newcommand{\vthree} {$v_{3}$\xspace}
\newcommand{\vn}     {$v_{n}$\xspace}
\newcommand{\pio}    {$\pi$}
\newcommand{\hyp}    {$^{3}_{\Lambda}\mathrm H$}
\newcommand{\antihyp}{$^{3}_{\bar{\Lambda}} \overline{\mathrm H}$}

\newcommand{\he}     {$^{3}\mathrm{He}$}
\newcommand{\antihe} {$^{3}\mathrm{\overline{He}}$}
\newcommand{\dedx}   {d$E$/d$x$}
\newcommand{\mom}    {\mbox{\rm MeV$\kern-0.15em /\kern-0.12em c$}}
\newcommand{\gmom}   {\mbox{\rm GeV$\kern-0.15em /\kern-0.12em c$}}
\newcommand{\mass}   {\mbox{\rm GeV$\kern-0.15em /\kern-0.12em c^2$}}
\newcommand{\Mmass}  {\mbox{\rm MeV$\kern-0.15em /\kern-0.12em c^2$}}
\newcommand{\Dp}     {$\Delta p_0$}
\newcommand{\Dr}     {$\Delta \rho_0$}

\newcommand{\red}{\textcolor{red}}

\begin{titlepage}
\PHyear{2020}       
\PHnumber{099}      
\PHdate{29 May}  


\title{Elliptic and triangular flow of \mbox{(anti)deuterons} in Pb--Pb collisions at $\sqrt{s_{\mathrm{NN}}}$~=~5.02~TeV}
\ShortTitle{Elliptic and triangular flow of \mbox{(anti)deuterons}}   

\Collaboration{ALICE Collaboration\thanks{See Appendix~\ref{app:collab} for the list of collaboration members}}
\ShortAuthor{ALICE Collaboration} 

\begin{abstract}

The measurements of the \mbox{(anti)deuterons} elliptic flow (\vtwo) and the first measurements of triangular flow (\vthree) in \PbPb collisions at a center-of-mass energy per nucleon--nucleon collisions \snn~= 5.02~TeV are presented. 
A mass ordering at low transverse momentum (\pt) is observed when comparing these measurements with those of other identified hadrons, as expected from relativistic hydrodynamics. 
The measured \mbox{(anti)deuterons} \vtwo lies between the predictions from the simple coalescence and blast-wave models, which provide a good description of the data only for more peripheral and for more central collisions, respectively. The mass number scaling, which is violated for \vtwo\, is approximately valid for the \mbox{(anti)deuterons} \vthree.
The measured \vtwo\ and \vthree\ are also compared with the predictions from a coalescence approach with phase-space distributions of nucleons generated by iEBE-VISHNU with AMPT initial conditions coupled with UrQMD, and from a dynamical model based on relativistic hydrodynamics coupled to the hadronic afterburner SMASH. The model predictions are consistent with the data within the uncertainties in mid-central collisions, while a deviation is observed in the most central collisions.
  
\end{abstract}
\end{titlepage}

\setcounter{page}{2} 


\section{Introduction}

The production mechanism of light (anti)nuclei in high-energy hadronic collisions is still not fully clear and is under intense debate in the scientific community~\cite{SHM1,SHM2,SHM5,iEBE_VISHNU,HydroSMASH}. The understanding of the production of loosely-bound multi-baryon states in heavy-ion collisions has additional complications due to the fact that the phase transition is followed by a hadrons gas phase with intense re-scattering of hadrons.
At the Large Hadron Collider (LHC) energies, the lifetime of the hadronic phase between chemical and kinetic freezeout is in the range \mbox{4--7 fm/$\textit{c}$}~\cite{LifetimeHadronicPhase} and the kinetic freezeout temperature, when elastic interactions cease, is of the order of 100~MeV \cite{SpectraIdentifiedHadrons276,pionsKaonsProtonsALICE5TeV}.
The binding energy of multi-baryon systems such as light \mbox{(anti)nuclei} typically does not exceed a few MeV, which is almost two orders of magnitude smaller than the temperature of the system. Considering the high density of hadrons in the post-hadronization stage and the large dissociation cross sections of light \mbox{(anti)nuclei}, it is not clear how such loosely-bound systems can survive under these extreme conditions. 

Existing phenomenological models provide very different interpretations for this observation. 
In the statistical hadronization model~\cite{SHM1,SHM2,SHM3,SHM4,SHM5}, light \mbox{(anti)nuclei} as well as all other hadron species are assumed to be emitted by a source in local thermal and hadrochemical equilibrium. Their abundances are fixed at the chemical freeze-out, occurring at a temperature of $T_{\mathrm{chem}}=156~\pm~4$~MeV for \PbPb\ collisions at the LHC~\cite{alpha}. This model provides a good description of the measured hadron yields in central nucleus--nucleus collisions~\cite{SHM1}. However, the mechanism of hadron production and the propagation of loosely-bound states through the hadron gas phase 
are not addressed by this model. In the context of the statistical hadronization model, it has been conjectured that such objects could be produced at the phase transition as compact colorless quark clusters with the same quantum numbers of the final state hadrons. The survival of these states at high temperatures is interpreted as due to the low interaction cross section with the surrounding medium~\cite{SHM1}. 

In the coalescence approach, multi-baryon states are assumed to be formed by the coalescence of baryons at the kinetic freeze-out. In the simplest  versions of this model~\cite{Coalescence1, Coalescence2}, baryons are treated as \mbox{point-like} particles and the coalescence happens instantaneously if the momentum difference between nucleons is smaller than a given threshold, which is typically of the order of 100~MeV/$c$, while spatial coordinates are ignored.
In the state-of-the-art implementations of the coalescence approach~\cite{Coalescence3, iEBE_VISHNU}, the \mbox{quantum-mechanical} properties of baryons and their bound states are taken into account and the coalescence probability is calculated from the overlap between the wave functions of baryons and the Wigner density of the final-state cluster. All light \mbox{(anti)nuclei} produced at the phase transition are assumed to be destroyed by the interactions in the hadron gas phase and regenerated  with the same amount only at the latest stage of the system evolution.

To address the open question of the survival of loosely-bound multi-baryon states in the hadron gas phase with intense re-scattering, models based on relativistic hydrodynamics coupled to a hadronic afterburner have been recently developed~\cite{iEBE_VISHNU,HydroSMASH}. 
In these models, nucleons and light nuclei are produced at the phase transition using the Cooper-Frye formula~\cite{CooperFrye}, which describes the hadron production based on the local energy density of the fireball, and their yields are fixed to the value predicted by the thermal model at the chemical freeze-out temperature. Their propagation through the hadronic medium is simulated based on known interaction cross sections and resonant states using different transport codes. Existing calculations are based on UrQMD~\cite{UrQMD1, UrQMD2}, with light nuclei being produced by nucleon coalescence, and SMASH~\cite{HydroSMASH}, where \mbox{(anti)deuterons} are assumed to be destroyed and regenerated with equal rates in the hadronic stage.
The model based on UrQMD with nucleon coalescence~\cite{iEBE_VISHNU} provides a good description of the elliptic flow of \mbox{(anti)deuterons} measured in \PbPb collisions at \snn~=~2.76~TeV~\cite{deuteronFlowALICE} and of that of \mbox{(anti)$^{3}$He} measured in \PbPb collisions at \snn~=~5.02~TeV~\cite{3HeFlowALICE}. The model is able to describe the low-\pt\ spectra of deuterons, but over-predicts the deuterons data above 2.5~\gmom\ and the \mbox{(anti)$^{3}$He} spectra in the full momentum interval.   
The hybrid model based on SMASH successfully describes the measured \mbox{(anti)deuterons} \pt-spectra and coalescence parameter $B_{2}$, defined as the ratio of the invariant yield of deuterons and that of protons squared, measured in \PbPb collisions at  \snn~=~2.76~TeV~\cite{deuteronFlowALICE}. 

A conceptually similar approach, based on the analogy between the evolution of the early universe after the Big Bang and the space--time evolution of the system created in heavy-ion collisions, has recently been developed~\cite{sahaEquation}. The production of light (anti)(hyper)nuclei in heavy-ion collisions at the LHC is considered in the framework of the Saha equation assuming that disintegration and regeneration reactions involving light nuclei proceed in relative chemical equilibrium after the chemical freeze-out of hadrons. 

The existing models depict radically different pictures of the post-hadronization stage for loosely-bound states. 
Considering this scenario, the measurements of radial and anisotropic flow of light \mbox{(anti)nuclei}, i.e. the harmonics ($v_{n}$) of the Fourier decomposition of their azimuthal production distribution with respect to a symmetry plane of the collision, are relevant to study their propagation through the hadron gas phase and the dynamics of their interactions with other particles. 
Compared to the elliptic flow, the triangular flow of light \mbox{(anti)nuclei} has a better sensitivity to the fluctuating initial conditions as well as the properties of the created systems. Therefore, tighter constrains to on theoretical model that describe the production mechanism of light \mbox{(anti)nuclei} can be set. 

The elliptic flow of \mbox{(anti)deuterons} was measured as a function of the transverse momentum (\pt) for different centrality classes in \PbPb collisions at \snn~=~2.76~TeV~\cite{deuteronFlowALICE}. A clear mass ordering is observed at low \pt (\pt~$<$~3~\gmom) when this measurement is compared to that of other hadrons species~\cite{FlowIdentifiedHadrons276}, as expected from relativistic hydrodynamics. 
The simple coalescence model, based on the assumption that the \mbox{(anti)deuterons} invariant yield is proportional to the invariant yield of \mbox{(anti)protons} squared, is found to overestimate the measured \vtwo\ in all centrality intervals. The data are better described by the blast-wave model, a simplified version of the relativistic hydrodynamic approach in which the collective expansion is described using a parameterized hydrodynamic flow field.
The elliptic flow of \mbox{(anti)$^{3}$He} was measured in \PbPb collisions at \snn~=~5.02~TeV~\cite{3HeFlowALICE}. Also in the case of \mbox{(anti)$^{3}$He}, the mass ordering is observed for \pt~$<$~3~\gmom\ and the measured elliptic flow lies between the predictions of the blast-wave~\cite{BlastWave1} and the simple coalescence model. A better description of the measurement is provided by a more sophisticated coalescence model where the phase-space distributions of protons and neutrons are generated by the iEBE-VISHNU hybrid model with AMPT initial conditions~\cite{iEBE_VISHNU}. 
The picture that has emerged so far, regarding the elliptic flow of (anti)nuclei measured at LHC energies, is that the simple coalescence and blast-wave models represent the upper and lower edges of a region where the data are mostly located. Recent developments in the coalescence approach, which take into account momentum-space correlations of nucleons and their quantum-mechanical properties, provide a better description of the data~\cite{iEBE_VISHNU, HydroSMASH}. 

In this paper, a precision measurement of the \mbox{(anti)deuterons} elliptic flow and first ever measurement of \mbox{(anti)deuterons} triangular flow for different \pt\ and centrality intervals in \PbPb collisions at \snn~= 5.02~TeV are presented. Thanks to the large data sample collected at higher energy, the elliptic flow measurement is performed in wider \pt and up to a higher centrality intervals compared to that in \PbPb collisions at \snn~=~2.76~TeV, allowing for a more differential comparison with the theoretical models.  

\section{The ALICE detector}
\label{sec:alice}
A detailed description of the ALICE detector can be found 
in~\cite{Abelev:2014ffa} and references therein.
The main sub-detectors used for  the present analysis are the V0 detector, 
the Inner Tracking System (ITS), the Time Projection Chamber (TPC), and the  
Time-of-Flight detector (TOF), which are located inside a solenoidal magnet that provides a uniform field of 0.5~T directed along the beam direction. 
The V0 detector~\cite{Abbas:2013taa} consists of two arrays of scintillation counters placed around the beam vacuum tube on either side of the interaction point: one covering the pseudorapidity interval $2.8 < \eta < 5.1$~\mbox{(V0A)} and the other one covering $-3.7 < \eta < -1.7$~\mbox{(V0C)}. Each V0 array consists of four rings in the radial direction, with each ring comprising eight cells with the same azimuthal size. The scintillator arrays have an intrinsic time resolution better than 0.5~ns, and their timing information is used in coincidence for offline rejection of events produced by the interaction of the beams with residual gas in the vacuum pipe. 
The V0 scintillators are used to determine the collision centrality from the measured charged-particle multiplicity~\cite{Aamodt:2011oai,Abelev:2013qoq,CentralityDeterminationPublicNote} 
and to measure the orientation of the symmetry plane of the collision.

The ITS~\cite{Aamodt:2010aa}, designed to provide high-resolution track points in the vicinity of the nominal vertex position, is composed of three subsystems of silicon detectors placed around the interaction region with a cylindrical symmetry. The Silicon Pixel Detector (SPD) is the subsystem closest to the beam vacuum tube and it is made of two layers of pixel detectors. The third and the fourth layers are formed by Silicon Drift Detectors (SDD), while the outermost two layers are equipped with double-sided Silicon Strip Detectors (SSD). 
The ITS covers the pseudorapidity interval $|\eta |<0.9$. 

The same pseudorapidity interval is covered by the TPC, which is the main tracking detector, consisting of a hollow cylinder whose axis coincides with the nominal beam axis. 

The active volume of 90 m$^{3}$ is filled with a gas mixture containing 88$\%$ Ar and 12$\%$ CO$_{2}$.

The trajectory of a charged particle is estimated using up to 159 space points.
The charged-particle tracks are then built by combining the hits in the ITS and the reconstructed space points in the TPC. 
The TPC is also used for particle identification (PID) by measuring the specific energy loss (\dedx) in the TPC gas.

The TOF detector~\cite{Akindinov:2013tea} covers the full azimuth in the 
pseudorapidity interval $|\eta|<0.9$. The detector is based on the Multi-gap Resistive Plate Chambers (MRPCs) technology and it is located, with a 
cylindrical symmetry, at an average radial distance of 380 cm from the
beam axis. The TOF allows for PID, based on the difference between the measured time-of-flight and its expected value, computed for each mass hypothesis from the track momentum and length. The resolution on the measurement of the time-of-flight is about 60~ps in heavy-ion collisions. 

\section{Data sample and analysis techniques}
\label{sec:Analysis}

\subsection{Event and track selections}
\label{subsec:EventAndTrackSelection}

The data sample used for the measurements presented in this paper was recorded by ALICE in 2015 during the LHC \PbPb run at \snn~=~5.02~TeV. A minimum bias trigger was used during the data taking, which required coincident signals in both V0 detectors. 
An offline event selection is applied to remove beam-gas collisions using the timing information provided by the V0 detectors and the Zero-Degree Calorimeters~\cite{Abelev:2014ffa}. 
Events with multiple primary vertices identified with the SPD are tagged as pileup and removed from the analysis.  
In addition, events with significantly different charged-particle multiplicities measured by the V0 detector and by the
tracking detectors at midrapidity, which have different readout times, are rejected. 
After the offline event selection, the remaining contribution of beam-gas events is smaller than 0.02\%~\cite{Abelev:2014ffa} and the fraction of pileup events is found to be negligible.
The primary vertex position is determined from tracks reconstructed in the ITS and TPC as described in~\cite{Abelev:2014ffa} and only events with a reconstructed primary vertex position along the beam axis within 10~cm from the from the nominal interaction point are selected.  
The total number of events selected for the analysis for centrality 0--70\% is about 73 million.

Deuteron (d) and antideuteron (\dbar) candidates are selected from charged-particle tracks reconstructed in the ITS and TPC in the
kinematic range $|\eta|~<~$0.8 and 0.8~$<$~\pt~$<$~6~GeV/$\textit{c}$. Only tracks with at least 70 clusters out of a maximum of 159 and with a $\chi^2$ per degree-of-freedom for the track fit lower than 2 are accepted. In addition, in order to guarantee a track-momentum resolution of $2\%$ in the measured \pt\ range and a \dedx\ resolution of about $6\%$, each track is required to be reconstructed from at least 80\% of the number of expected TPC clusters and to have at least one hit in either of the two innermost layers of the ITS. 
The distances of closest approach (DCA) to the primary vertex in the plane perpendicular and parallel to the beam axis for the selected tracks are determined with a resolution better than 300~$\mu$m~\cite{Abelev:2014ffa}. To suppress the contribution of secondary particles, the reconstructed tracks are required to have a longitudinal DCA smaller than 2~cm and a transverse DCA smaller than $0.0105+0.0350/p_{\mathrm{T}}^{1.1}$~cm, with $p_{\mathrm{T}}$ in units of $\mathrm{GeV}/\textit{c}$. The latter corresponds to approximately $7\sigma_{\rm DCA} (p_{\rm T})$, where $\sigma_{\rm DCA}(p_{\rm T})$ is the transverse DCA resolution in the corresponding \pt interval.

\subsection{\mbox{(Anti)deuterons} identification}
\label{subsec:PID}

The \mbox{(anti)deuterons} identification technique used in this analysis is similar to that used in the previous measurement in \PbPb collisions at \snn~=~2.76~TeV~\cite{deuteronFlowALICE}. 
For transverse momenta up to 1.4~\gmom\, \mbox{(anti)deuterons} are identified using the only the TPC information by requiring that the average \dEdx\ is within 3$\sigma$ from the expected average value for the \mbox{(anti)deuteron} mass hypothesis. 
For \pt~$>$~1.4~\gmom\, the 3$\sigma$ TPC identification is complemented by the signal provided by the TOF detector. The number of \mbox{(anti)deuterons} in each \pt interval is extracted from a fit of the $\Delta$M = $m_{\mathrm{TOF}} - m_{\mathrm{d_{pdg}}}$, where $m_{\mathrm{TOF}}$ is the particle mass calculated using the time-of-flight measured by the TOF and  $m_{\mathrm{d_{pdg}}}$ is the nominal mass of deuterons taken from~\cite{PDGdeuteron}. In the left panel of Fig.~\ref{fig:method} the  $\Delta$M distribution for \mbox{(anti)deuterons} with $2.2 \leq$~\pt~$<2.4$~\GeVc in the centrality interval 20--30\%, is shown. The d+\dbar\ signal is fitted using a Gaussian with an exponential tail, while the background, originating from TOF hits incorrectly associated to tracks extrapolated from the TPC, is modeled with an exponential function. 

Deuterons and antideuterons are summed together (\dbar+d) in all the centrality intervals and for \pt larger than 1.4~\gmom. This is possible since the \vtwo and \vthree measured for \vtwo and \vthree for d and \dbar\ are consistent within the statistical uncertainties. 
At lower \pt, deuterons produced by spallation in interactions between  particles and the detector material or in the beam vacuum tube constitute a  significant background. For this reason, for \pt$<$~1.4~\gmom\ only antideuterons, which are not affected by this background, are used in the analysis. Since no difference is expected for the \vtwo\ and \vthree\ of \dbar\ and d, hereafter deuterons will denote results for antideuteron for \pt~$<$~1.4~\gmom\ and the sum of d and \dbar\ elsewhere. 
The contribution of secondary deuterons produced in weak decays of hypertritons is negligible considering that the production rate of \mbox{(hyper)nuclei} with mass number $A=3$ is suppressed compared to that of $A=2$ by a factor of approximately 300 in \PbPb collisions at \snn~=~2.76~TeV~\cite{Adam:2015vda}. 
A similar suppression is expected in \PbPb collisions at \snn~=~5.02~TeV.

\subsection{Flow analysis techniques}
\label{subsec:flow}

The particle azimuthal distribution of charged particles with respect to the \textit{n}-th order flow  symmetry plane $\Psi_{n}$~\cite{Voloshin:2008dg,Ollitrault:2009ie,Alver:2010gr,Qiu:2011iv} can be expressed as a Fourier series 
 \begin{equation}
 E \frac{\mathrm{ d^3} N}{\mathrm{d} p^3} = \frac{1}{2\pi} \frac{\mathrm {d^2} N}{p_\mathrm{T} \mathrm{d} p_{\rm{T}} \mathrm{d} y } \left( 1 + \sum_{n=1}^{\infty} 2 v_n \cos \left( n \left( \varphi -  \Psi_{n} \right) \right) \right),
 \end{equation}
 where $E$ is the energy of the particle, $p$ the momentum, $\varphi$ the azimuthal angle, $y$ the rapidity, 
 and  
 \begin{equation}
 v_n = \langle \cos \left( n(\varphi - \Psi_{n}) \right)\rangle .
 \end{equation}

 The second coefficient of the Fourier series (\vtwo) is called elliptic flow and is related to the initial geometrical anisotropy of the overlap region of the colliding nuclei. The third-order flow coefficient (\vthree), called triangular flow, is generated by fluctuations in the initial distribution of nucleons and gluons  in the overlap region~\cite{Bhalerao:2006tp, Alver:2010gr, Alver:2010dn}. The same fluctuations are responsible for the \vtwo\ measured in most central  collisions (centrality $<$~5\%)~\cite{ALICE:2011ab}. 
The $v_n$ coefficients are measured using the Scalar Product (SP) method~\cite{Adler:2002pu,Voloshin:2008dg}. This is a two-particle correlation technique based on the scalar product of the unit flow vector of the particle of interest, $k$, and the Q-vector. The unit flow vector is denoted by \textbf{u}$_{n,k} = \mathrm{exp}(in\varphi_k)$,  where $\varphi_k$ is the azimuthal angle of the particle $k$. 

The Q-vector is computed from a set of reference flow particles and is defined as: 
\begin{equation}
    \textbf{Q}_n = \sum w_i e^{in \varphi_i}
\end{equation}
where, in general, $\varphi_i$ is the azimuthal angle for the i-th reference flow particle, \textit{n} is the order of the harmonic, and $w_i$ is a weight applied to correct for reference flow. 

The $v_n$ flow coefficients are calculated as
\begin{equation}
    v_n \{\mathrm{SP}\} = \frac{\left\langle \langle \textbf{u}_{n,k}{\textbf{Q}^*_n} \rangle \right\rangle}{\sqrt{\frac{\langle \textbf{Q}_n \textbf{Q}^{A*}_n \rangle \langle \textbf{Q}_n \textbf{Q}^{B*}_n \rangle} {\langle \textbf{Q}^A_n \textbf{Q}^{B*}_n \rangle}}} .
\end{equation}

Single brackets $\langle ... \rangle$ denote an average over all events, while double brackets $\langle \langle ... \rangle \rangle$ indicate an average over all particles in all events, and ${}^*$ denotes the complex conjugate. The denominator is a correction factor that is introduced to take into account the resolution of the $\textbf{Q}_{n}$ vector. In this analysis, the  $\textbf{Q}_{n}$ vector is calculated from the azimuthal distribution of the energy deposition measured in the V0A, while the $\textbf{Q}_{n}^A$ and $\textbf{Q}_{n}^B$ vectors  are determined from the azimuthal distribution of the energy deposited in the V0C and the azimuthal distribution  of tracks reconstructed in the TPC, respectively. Using these detectors, a pseudorapidity gap $|\Delta \eta| > 2$ between the particle of interest and the reference flow particles is introduced. Such a pseudorapidity gap reduces non-flow effects, which are correlations not arising from the collective expansion of the system (e.g. resonances decays and jets). 

\begin{figure}[!htb]
\begin{tabular}{ccc}
\begin{minipage}{.3\textwidth}
\centerline{\includegraphics[width=1\textwidth]{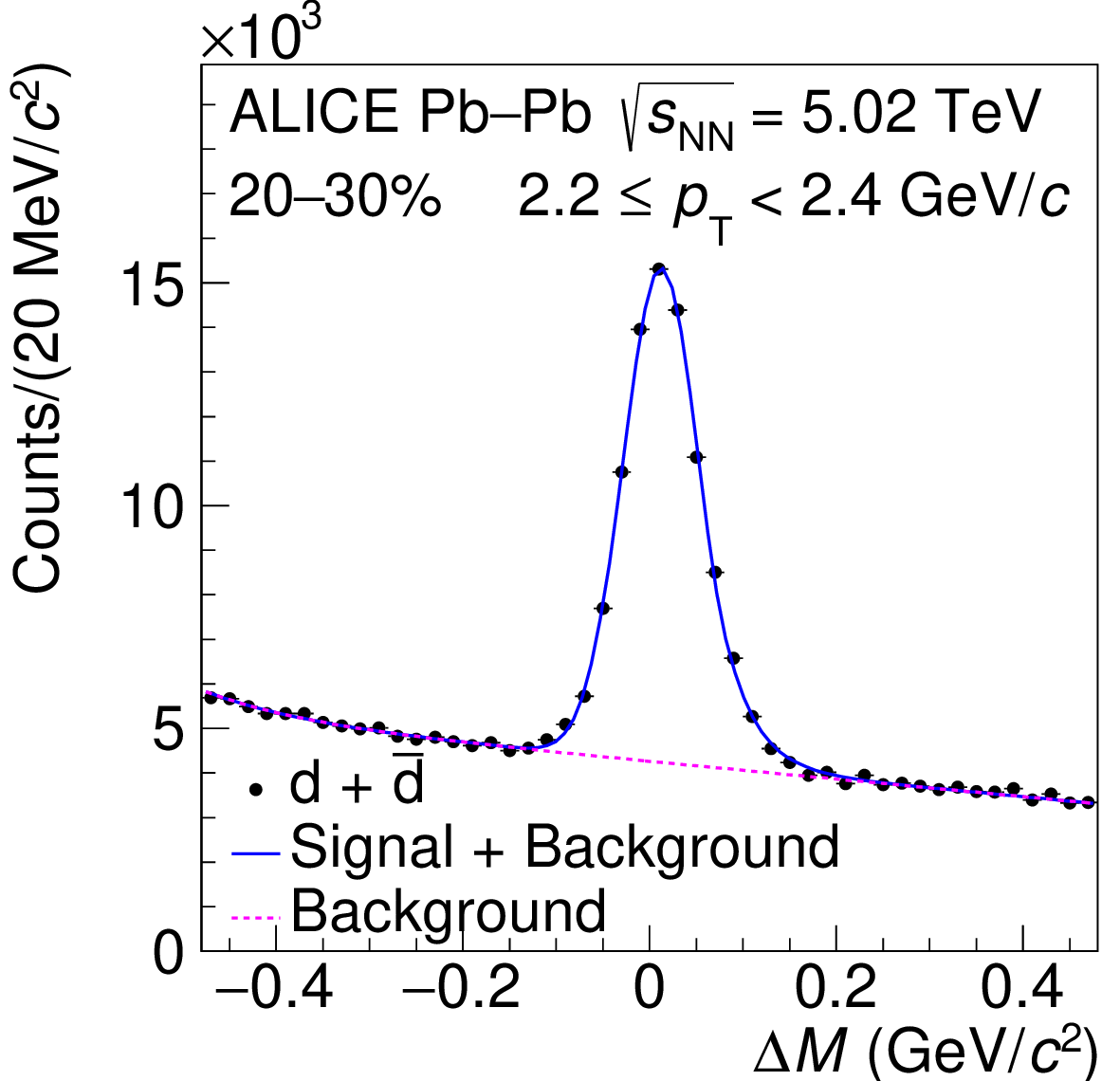}}
\end{minipage} & 
\begin{minipage}{.3\textwidth}
\centerline{\includegraphics[width=1\textwidth]{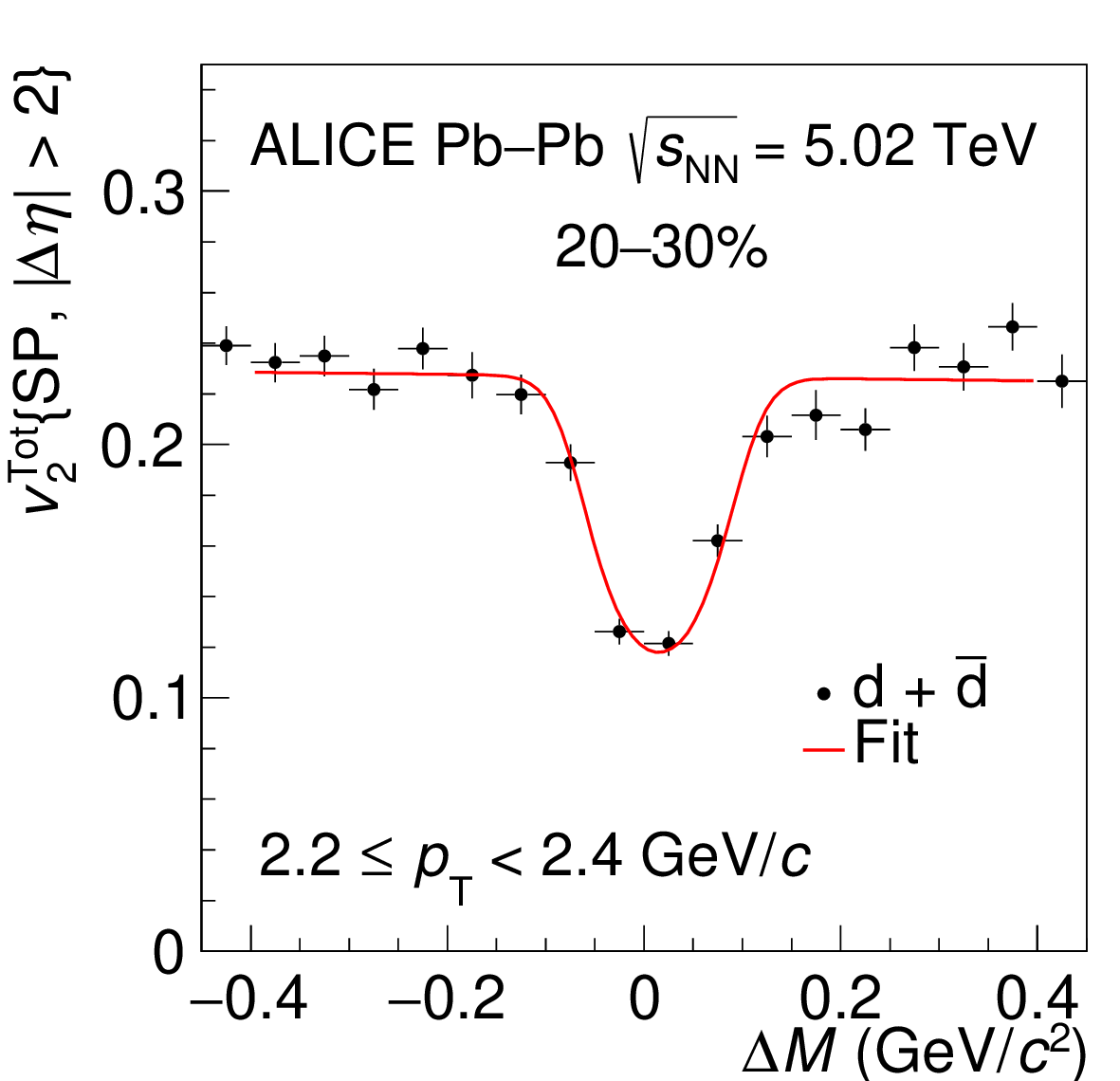}}
\end{minipage} & 
\begin{minipage}{.3\textwidth}
\centerline{\includegraphics[width=1\textwidth]{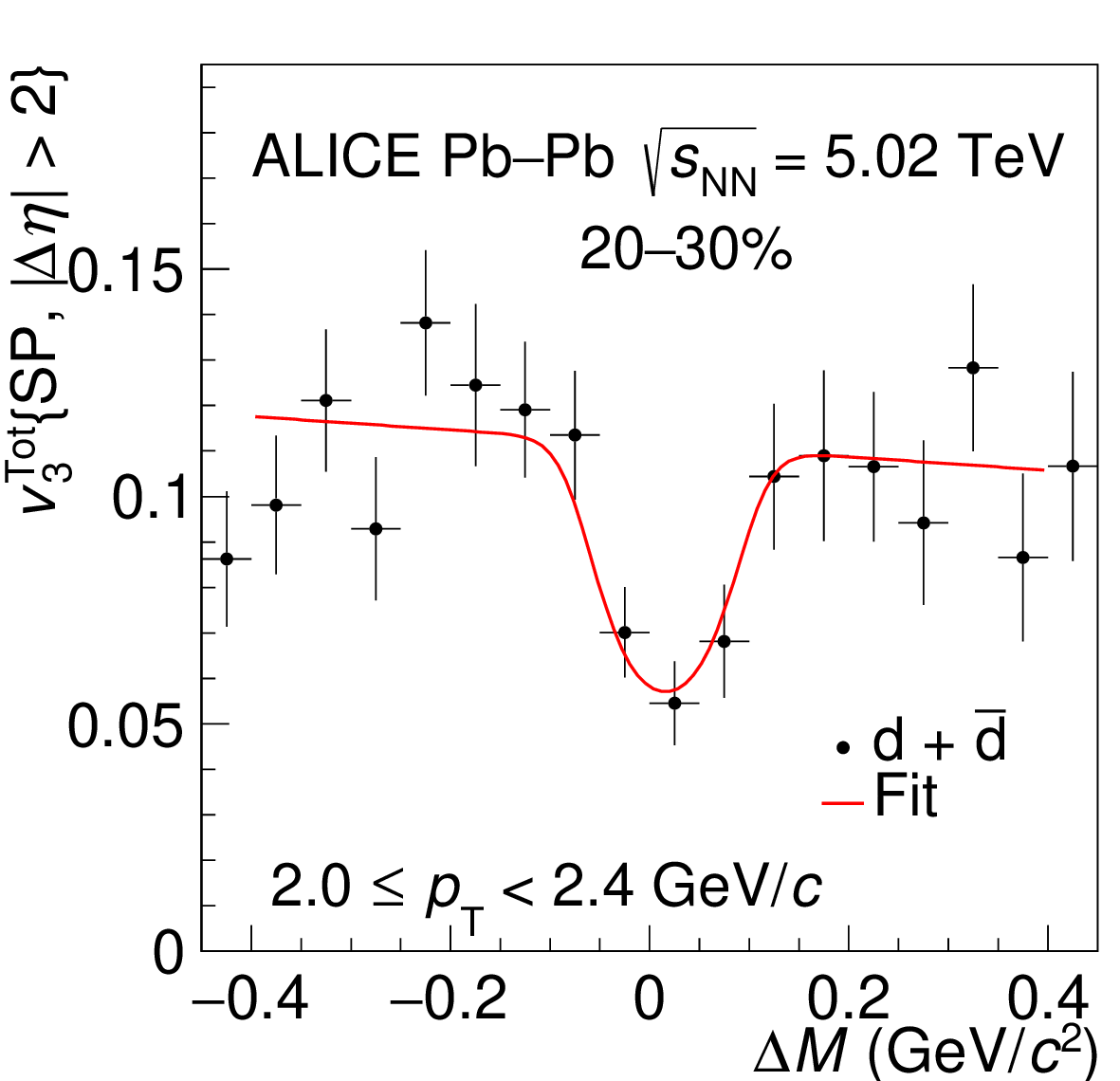}}
\end{minipage} 
\end{tabular}
\caption{(Color online) Raw yield (left), \vtwo (middle) and \vthree (right) of d+\dbar\ candidates as a function of $\Delta \rm{M}$ for $2.2~\leq$~\pt~$<2.4$~\GeVc ($2.0~\leq$~\pt~$<2.4$~\GeVc for \vthree) and in the centrality interval 20--30\%. The data points represent the measurements. The curve on the left panel is the total fit (signal
plus background) as described in the text. The curves in the middle and right panel are the fits performed using Eq.~\ref{eq:v2tot}. Vertical bars represent the statistical uncertainties.}
\label{fig:method}
\end{figure}

The purity of the sample of deuterons identified using the TPC in the 0.8~$<$~\pt~$<$~1.4~\gmom\ interval is around 100\%. In this transverse momentum interval the \vtwo\ and \vthree\ coefficients were evaluated on a track-by-track basis and then averaged in each \pt\ interval. 
For higher \pt, the $v_{n}$ coefficients are calculated in different ranges of $\Delta$M. 
The $v_{n}$($\Delta$M) contains contributions from the signal ($v_n^{\mathrm{sig}}$) and from the background ($v_n^{\mathrm{bkg}}$)

\begin{equation} 
v_n (\Delta\mathrm{M}) = v_n^{\mathrm{sig}} \frac{\mathrm{N}^{\mathrm{sig}}}{\mathrm{N}^{\mathrm{tot}}}(\Delta\mathrm{M}) +  v_n^{\mathrm{bkg}}(\Delta\mathrm{M}) \frac{\mathrm{N}^{\mathrm{bkg}}}{\mathrm{N}^{\mathrm{tot}}}(\Delta\mathrm{M}),
\label{eq:v2tot}
\end{equation}

where N$^{\rm{sig}}$ is the number of \mbox{deuterons}, N$^{\rm{bkg}}$ the number of background particles and N$^{\rm{tot}}$ is their sum. 
The signal $v_{n}$ is extracted from a fit to the observed $v_{n}$ as a function of $\Delta$M, in which $v_n^{\mathrm{bkg}}$ is described using a first-order polynomial function, and $v_n^{\mathrm{sig}}$ is a free fit parameter. N$^{\rm{sig}}$ and N$^{\rm{bkg}}$ are obtained from the fit to the $\Delta \rm{M}$ distribution using a Gaussian with an exponential tail for the signal and an exponential for the background. 
The signal extraction procedure is illustrated in Fig.~\ref{fig:method} for 2.2~$\leq~$\pt~$<$~2.4 \GeVc (2.0~$\leq~$\pt~$<$~2.4 \GeVc for \vthree) in the centrality interval 20--30$\%$.

The elliptic and triangular flow of \mbox{deuterons} are measured in centrality intervals of 5$\%$ width and then the results in wider centrality intervals are obtained as weighted averages of these measurements using the number of deuteron candidates, in the same centrality interval of 5$\%$ width as a weight, similarly to what was performed in~\cite{3HeFlowALICE}. 

\subsection{Systematic uncertainties}
\label{subsec:SystematicUncertainties}

The sources of systematic uncertainties for the elliptic and triangular flow of deuterons are related to event selection, tracking, (anti-)deuterons identification, and the technique used for the signal extraction. 
The contribution related to the event selection is estimated by taking into account the differences in the $v_{2}$ and $v_{3}$ measurements obtained using different event-selection criteria. In particular, the fiducial region for the vertex position along the beam axis is varied from the range $[-10,10]$ cm to $[-7,7]$ cm to probe the magnitude of potential edge effects. 
To investigate possible effects due to charge asymmetries during tracking and geometrical asymmetries in the detector, the differences between the results obtained by using opposite magnetic field polarities are included. Analogously, the default centrality estimator is changed to that based on the number of hits in the first or second layer of the ITS. 
Finally, the effect related to pileup rejection is tested by requiring a stronger correlation between the V0 and central barrel multiplicities. These contributions are assumed to be independent and added in quadrature. The total systematic uncertainty due to event selection is found to be around 1.5$\%$ for both $v_{2}$ and $v_{3}$. 

To estimate the systematic uncertainties due to reconstruction and identification of deuterons, the track selection and the TPC PID criteria are varied with respect to the default choice and the $v_{n}$ measurements are repeated for each of these different settings. The RMS of the distribution of $v_{n}$ measurements in each \pt interval is considered as systematic uncertainty. To minimize the effect of statistical fluctuations, all variations smaller than 2$\sqrt{\left|\sigma_{0}^{2} - \sigma_{i}^{2}\right|}$ are not included in the estimate of the systematic uncertainties~\cite{Barlow:2002yb}, where $\sigma_{0}$ is the statistical uncertainty of the default value while $\sigma_{i}$ is that corresponding to the~$i^{\rm th}$ selection criterion. 
The probability distribution for the variations of data points due to systematic effects related to tracking and PID is assumed to be uniform in each \pt interval and the difference between the maximum and minimum value divided by $\sqrt{12}$ is assigned as systematic uncertainty. This contribution ranges from 1$\%$ and 3$\%$ depending on \pt and centrality. 

To estimate the contribution to the systematic uncertainties due to the signal extraction, the function used to describe the $v_n^{\mathrm{bkg}}$ is changed. In addition to a first-order polynomial, a constant function and a second-order polynomial are also used, and the maximum difference with respect to the default measurement is considered as systematic uncertainty. A contribution up to 5\% is observed for central collisions and for \pt~$<$~2~\gmom. Moreover, different functions and fitting ranges are used to describe the signal and the background of Eq.~\ref{eq:v2tot}. More specifically, besides a Gaussian function with an exponential tail, a Gaussian is also used for the signal, while single and double exponential, and linear functions are also used for the background. 
This contribution is relevant only for \pt $>$ 1.4~\gmom, where the TOF is used to extract the signal, and is found to vary from 1\% to 6\% depending on \pt and centrality. 
Table~\ref{tab:SystematicUncertaintiesTable} shows the summary of the different contributions to the systematic uncertainties for the \vtwo\ and \vthree\ of deuterons. The total uncertainties are given by their sum in quadrature, assuming that all contributions are independent. 
 
\begin{table}[!hbt]
\begin{center}
\caption {Summary of the systematic uncertainties for the deuterons \vtwo\ and \vthree. The maximum deviation of the systematic uncertainty is reported.}
\centering
\begin{tabular}{lcc}
\hline
\hline  Source                                                         & \multicolumn{2}{c} {Value}    \\
\hline                                                                 & \vtwo    & \vthree\\
\hline  
Event selections                     &   1.5\%   &  1.5\%  \\
Tracking and particle identification &  1--3\% & 1--2\% \\
Signal extraction                    &  1--4\% & 2--6\% \\ 
Total                                &  2--7\% & 3--7\% \\      
\hline
\end{tabular} 
\label{tab:SystematicUncertaintiesTable}

\end{center}
\end{table}

\section{Results and discussion}
\label{section:comparisons}

\begin{figure}[!hbt]
    \begin{center}
    \includegraphics[width = 0.49\textwidth]{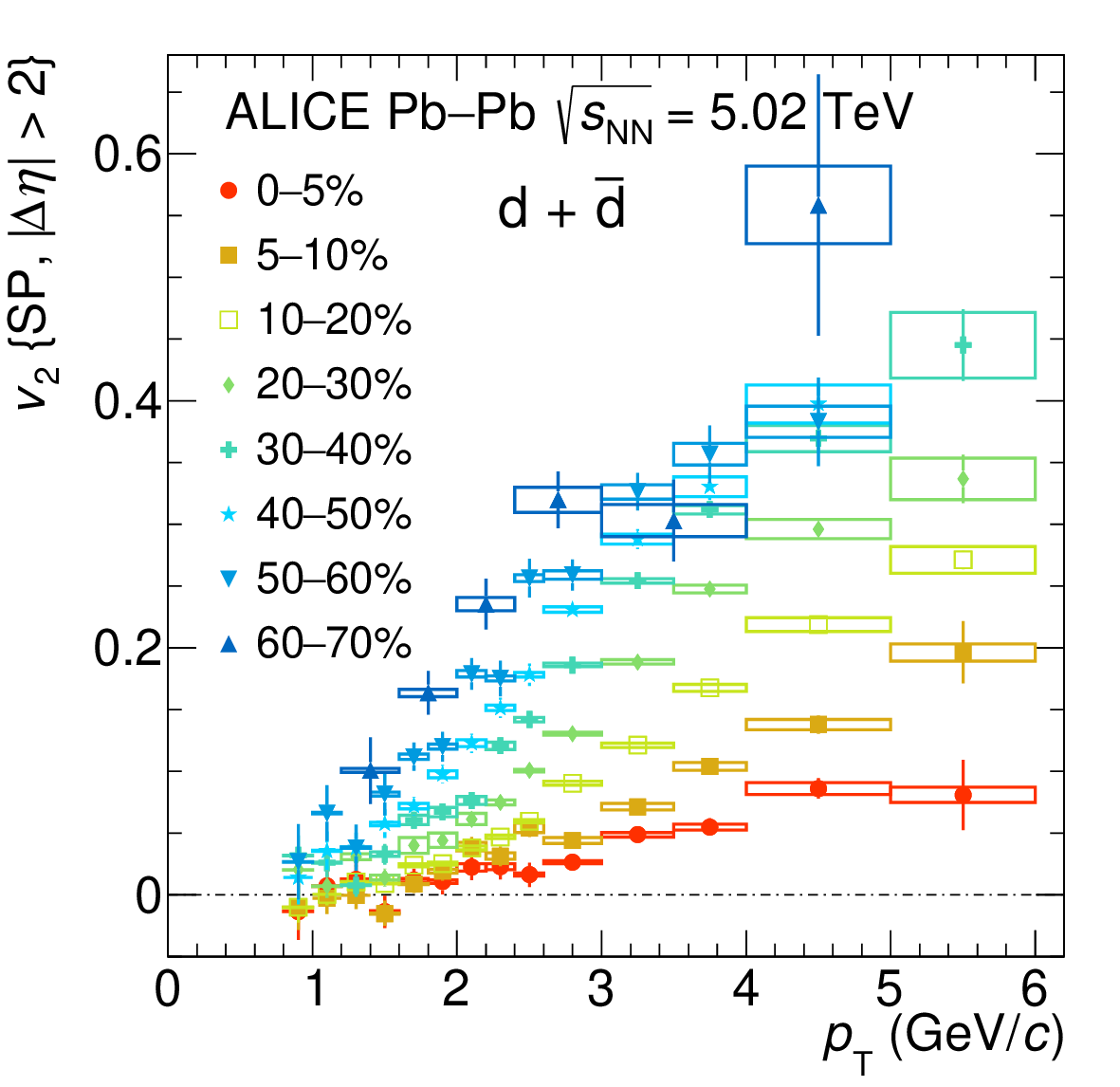}
    \includegraphics[width = 0.49\textwidth]{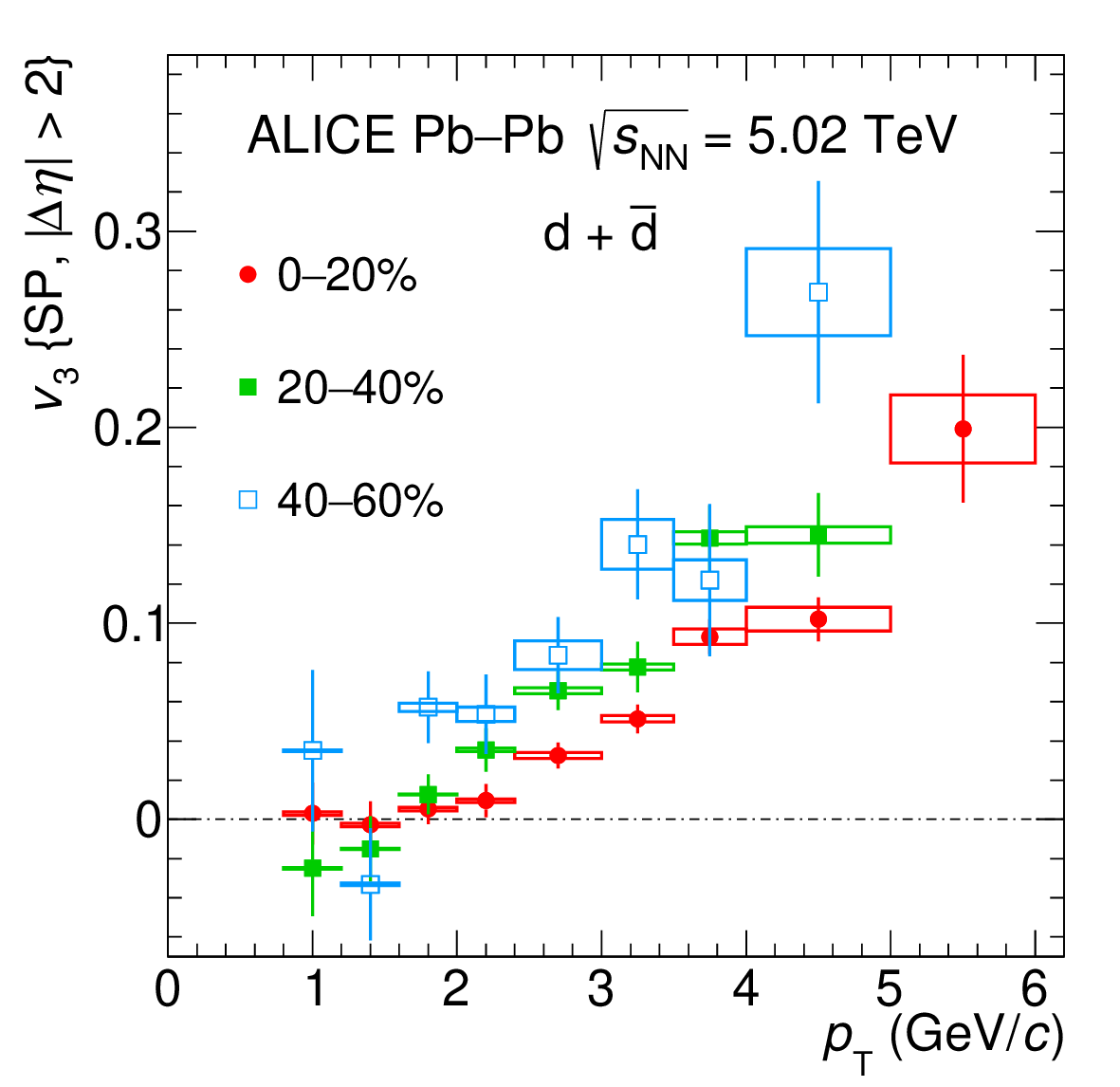}
    \end{center}
    \caption{(Color online) Elliptic (\vtwo, left) and triangular (\vthree, right) flow of \mbox{deuterons} as a function of \pt for different centrality intervals measured in \PbPb collisions at \snn~=~5.02~TeV. The horizontal line at zero is to guide the eye. Vertical bars and boxes represent the statistical and systematic uncertainties, respectively.}
\label{fig:Measurement}
\end{figure}

The \vtwo\ and \vthree\ of deuterons measured in \PbPb collisions at \snn~=~5.02~TeV are shown in Fig.~\ref{fig:Measurement} as a function of \pt for different centrality intervals. In the measured \pt\ interval, an increasing trend is observed with increasing \pt and going from central to more peripheral \PbPb collisions, as expected based on the relativistic hydrodynamic description of the collective expansion of a hot and dense medium~\cite{relativisticHydro}. Initial state fluctuations of the energy density distribution of partons in the colliding nuclei imply a non-zero \vthree~\cite{Alver:2010dn}.

The measurement presented in this paper shows that these initial state effects, already observed for other hadron species at LHC energies~\cite{Adam:2016nfo,identifiedHadronsFlow}, are also visible for \mbox{deuterons}.

\begin{figure}[!htb]
\begin{center}
\includegraphics[width=0.9\textwidth]{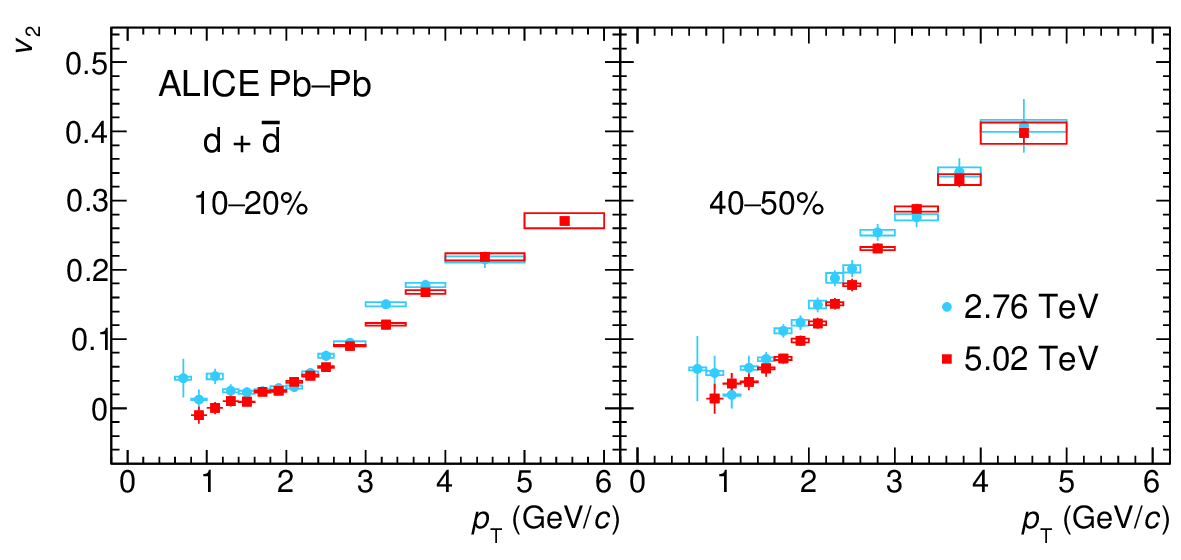}
\caption{(Color online) Deuterons \vtwo\ measured in \PbPb collisions at \snn~=~5.02~TeV (red square) compared to that measured at  \snn~=~2.76~TeV~\cite{deuteronFlowALICE} (light blue circles) for two centrality intervals (10--20\%  and 40--50\%). Both protons and deuteron elliptic flow were measured for pseudorapidity gap between the particle of interest and the reference flow particle  $|\Delta \eta| >$~0.9. 
Vertical bars and boxes represent the statistical and systematic uncertainties, respectively.}
\label{fig:run1run2direct}
\end{center}
\end{figure}

The measurement of the deuterons \vtwo in \PbPb collisions at \snn~=~5.02~TeV is compared to that in \PbPb collisions at \snn~=~2.76~TeV~\cite{deuteronFlowALICE} in Fig.~\ref{fig:run1run2direct} for two centrality intervals. The observed \vtwo and their trend are similar at the two center-of-mass energies, but a decrease of the observed elliptic flow for a given~\pt is observed with increasing center-of-mass energy. This effect is more pronounced in peripheral rather than in central collisions. A similar effect was observed for the \mbox{proton} \vtwo measurements~\cite{identifiedHadronsFlow} and is interpreted as partially due to the increasing radial flow with increasing collision energy, which produces a shift of the \vtwo\ towards higher \pt. 

\begin{figure}[!htb]
\begin{center}
\includegraphics[width=0.9\textwidth]{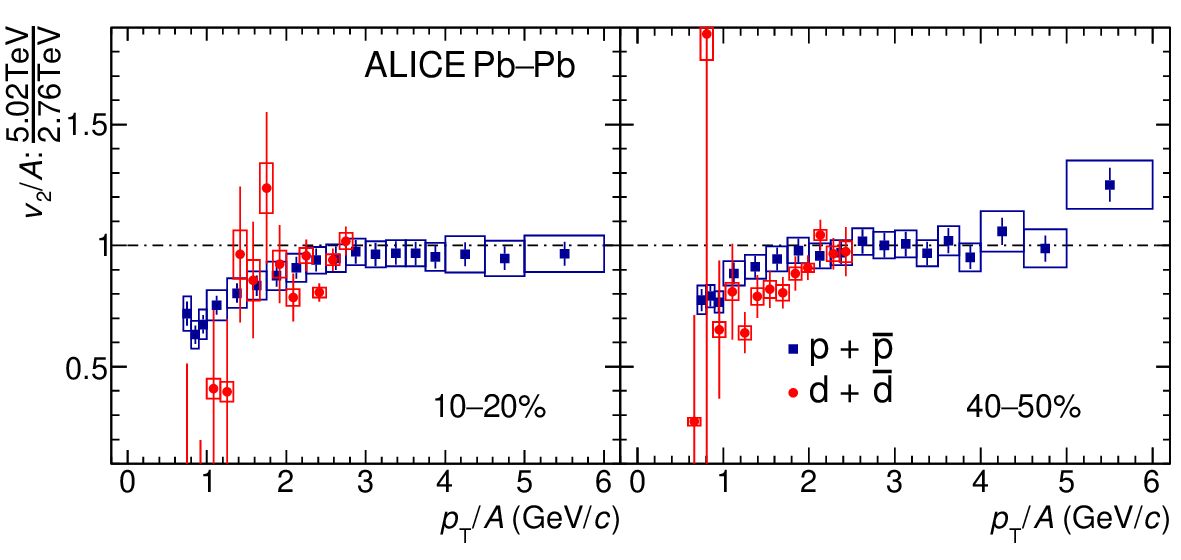}
\caption{(Color online) Ratio of the \vtwo\ of deuterons measured in \PbPb collisions at \snn~=~5.02~TeV to that measured at  \snn~=~2.76~TeV (red circles) compared with the same ratio obtained for protons (blue squares) for two centrality intervals (10--20\% on the left panel and 40--50\% on the right panel). For a direct comparison of protons and deuterons, the measured \vtwo\ and \pt\ have were divided by $A$. Vertical bars and boxes represent the statistical and systematic uncertainties, respectively.}
\label{fig:run1run2}
\end{center}
\end{figure}

The effect due to radial flow is assessed quantitatively by comparing the ratio of the deuteron and proton \vtwo as a function of \pt\ at the two energies. The ratio between the deuteron \vtwo in \PbPb collisions at \snn~=~5.02~TeV to that measured at \snn~=~2.76~TeV, with \vtwo\ and \pt\ scaled by the mass number $A = 2$, is shown in Fig.~\ref{fig:run1run2} for two centrality intervals in comparison with the same ratio for protons. As indicated by these ratios, the radial flow effects are quantitatively very similar for protons and deuterons. 
It has to be noted that a mass scaling would lead to the same conclusion since the binding energy of deuteron is of 2.2~MeV, i.e. the deuteron mass is approximately equal to 2$m_{\mathrm{p}}$, where $m_{\mathrm{p}}$ is the proton mass.

\begin{figure}[!htb]
\begin{center}
\includegraphics[width=1\textwidth]{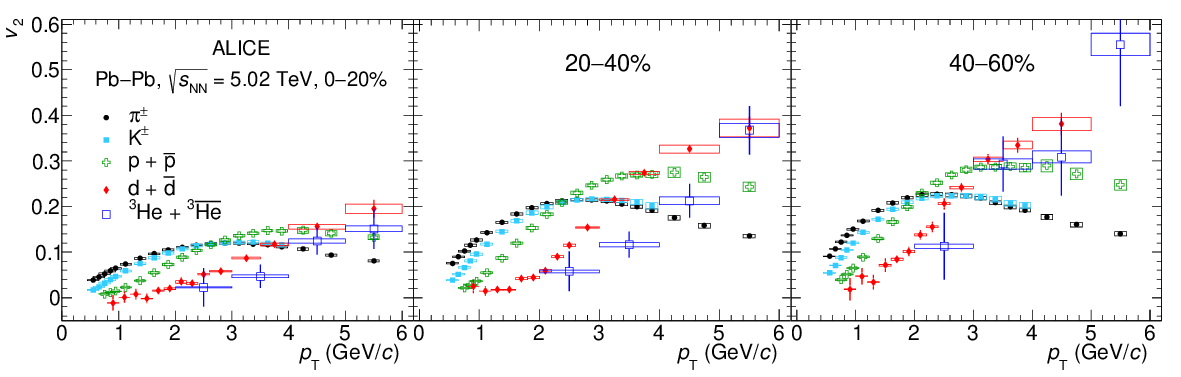}
\caption{(Color online) Comparison of the elliptic flow of pions, kaons, protons, deuterons and (anti)$^{3}$He in different centrality intervals for \PbPb\ collisions at \snn~=~5.02~TeV. (Anti)$^{3}$He \vtwo\ is measured using the Event Plane method~\cite{3HeFlowALICE}. Vertical bars and boxes represent the statistical and systematic uncertainties, respectively.}
\label{fig:ComparisonV2}
\end{center}
\end{figure}

The elliptic flow of deuterons is compared to that of pions, kaons, protons and \mbox{(anti)}$^{3}\mathrm{He}$ measured at the same center-of-mass energy~\cite{identifiedHadronsFlow,3HeFlowALICE}  in Fig.~\ref{fig:ComparisonV2}. Since the (anti)$^{3}$He elliptic flow is measured in centrality intervals of $20\%$ width due to its rarer production compared to that of lighter hadrons, the \vtwo of pions, kaons, protons and deuterons are re-calculated to match the same centrality intervals. This is achieved by averaging the \vtwo measurements of these particles in narrower centrality intervals weighted by the corresponding \pt spectra~\cite{pionsKaonsProtonsALICE5TeV,ALICE-PUBLIC-2017-006}. A clear mass ordering of \vtwo is observed at low \pt, as expected for a system expansion driven by the pressure gradient as described by relativistic hydrodynamics ~\cite{Huovinen:2001cy, Shen:2011eg,relativisticHydro}.

\begin{figure}[!htb]
\begin{center}
\includegraphics[width=0.8\textwidth]{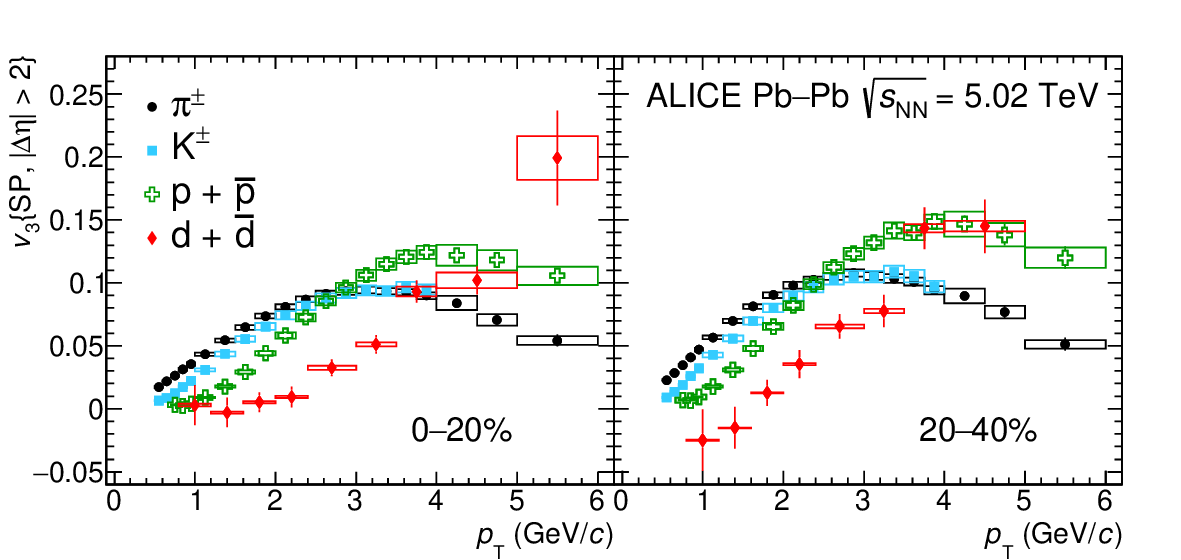}
\caption{(Color online) Triangular flow (\vthree) of deuterons, pions, kaons, and protons~\cite{identifiedHadronsFlow} as a function of \pt for the centrality intervals 0--20$\%$ and 20--40$\%$. Vertical bars and boxes represent the statistical and systematic uncertainties, respectively.}
\label{fig:ComparisonV3}
\end{center}
\end{figure}

In  Fig.~\ref{fig:ComparisonV3}, the deuterons \vthree\ is compared to that of pions, kaons, and protons  at the same center-of-mass energy~\cite{identifiedHadronsFlow} for the centrality intervals 0--20$\%$ (left) and 20--40$\%$ (right). Also for the \vthree, a clear mass ordering is observed for \pt~$\lesssim$~2.5~\gmom\ and \pt~$\lesssim$~3~\gmom\ for the centrality intervals 0--20$\%$ and 20--40$\%$, respectively. 

\subsection{Comparison with the blast-wave model predictions}
\label{subsec:BlastWaveComparison}

The elliptic flow of deuterons is compared with the expectations of the blast-wave model\mbox{~\cite{BlastWave1,BlastWave2,BlastWave3}}, which is 
based on the assumption that the system produced in heavy-ion collisions is locally thermalized and expands collectively with a common velocity field. The system is assumed to undergo an instantaneous kinetic freeze-out at the temperature $T_{\mathrm{kin}}$ and to be characterized by a common transverse radial flow velocity at the freeze-out surface.
A simultaneous fit of the $v_{2}$ and the \pt\  spectra of pions, kaons, and protons~\cite{identifiedHadronsFlow,pionsKaonsProtonsALICE5TeV} with  the blast-wave model is performed  in the transverse-momentum ranges 
0.5~$\leq$~$p^{\pi}_{\mathrm{T}}$~$<$~1~\gmom, 
$0.7 \leq p^{\mathrm{K}}_{\mathrm{T}} < 2$~\gmom, and  $0.7 \leq p^{\mathrm{p}}_{\mathrm{T}} < 2.5$~\gmom. 
The four free parameters of the blast-wave function are the kinetic freeze-out temperature ($T_{\mathrm{kin}}$), the variation in the azimuthal density of the source ($s_{2}$), the mean transverse expansion rapidity ($\rho_{0}$), and the amplitude of its azimuthal variation ($\rho_{\mathrm{a}}$), as described in~\cite{BlastWave2}. The values of these parameters extracted from the fits are reported in \autoref{Table:BlastWaveParameters} for each centrality interval. These values are employed to predict the elliptic flow of deuterons under the assumption that the same kinetic freeze-out conditions apply for all particles produced in the collision. The deuteron mass is taken from~\cite{PDGdeuteron}.

\begin{table}[!hbt]
\begin{center}
\centering
\renewcommand{\arraystretch}{1.2}
\caption{Blast-wave parameters extracted from the simultaneous fits of the \pt spectra and $v_{2}$ of pions, kaons, and protons measured at \snn~=~5.02~TeV. See text for details. The error assigned to each parameter is shown only with one significant digit.}
\begin{tabular}{ccccc}
\hline
Centrality &  \multicolumn{4}{c}{Fit parameters}\\
\hline
   &   $T_{\mathrm{kin}}$ (MeV)           &     $s_{2}$  ($10^{-2}$)      &             $\rho_{0}$ ($10^{-1}$)           &     $\rho_{\mathrm{a}}$ ($10^{-2}$) \\
\hline
       0--5\% & 104~$\pm$~1 & 2.63~$\pm$~0.01 & 8.57~$\pm$~0.01 & 0.83 ~$\pm$~0.01 \\
 5--10\% & 106~$\pm$~1 & 4.15~$\pm$~0.01 & 8.85~$\pm$~0.01 & 1.47 ~$\pm$~0.01 \\
10--20\% & 107~$\pm$~1 & 6.09~$\pm$~0.01 & 9.12~$\pm$~0.01 & 2.17 ~$\pm$~0.01 \\
20--30\% & 109~$\pm$~1 & 8.25~$\pm$~0.01 & 9.02~$\pm$~0.01 & 2.85 ~$\pm$~0.01 \\
30--40\% & 111~$\pm$~1 & 10.1~$\pm$~0.01 & 8.61~$\pm$~0.01 & 3.25 ~$\pm$~0.01 \\
40--50\% & 116~$\pm$~1 & 12.3~$\pm$~0.01 & 7.73~$\pm$~0.01 & 3.30 ~$\pm$~0.01 \\
50--60\% & 121~$\pm$~1 & 14.5~$\pm$~0.01 & 6.93~$\pm$~0.01 & 2.85 ~$\pm$~0.01 \\
60--70\% & 129~$\pm$~1 & 17.4~$\pm$~0.01 & 5.95~$\pm$~0.01 & 1.74 ~$\pm$~0.01 \\
     
\hline
\end{tabular}
\label{Table:BlastWaveParameters}
\end{center}
\end{table} 

\begin{figure}[!htb]
\begin{center}
\includegraphics[width=0.8\textwidth]{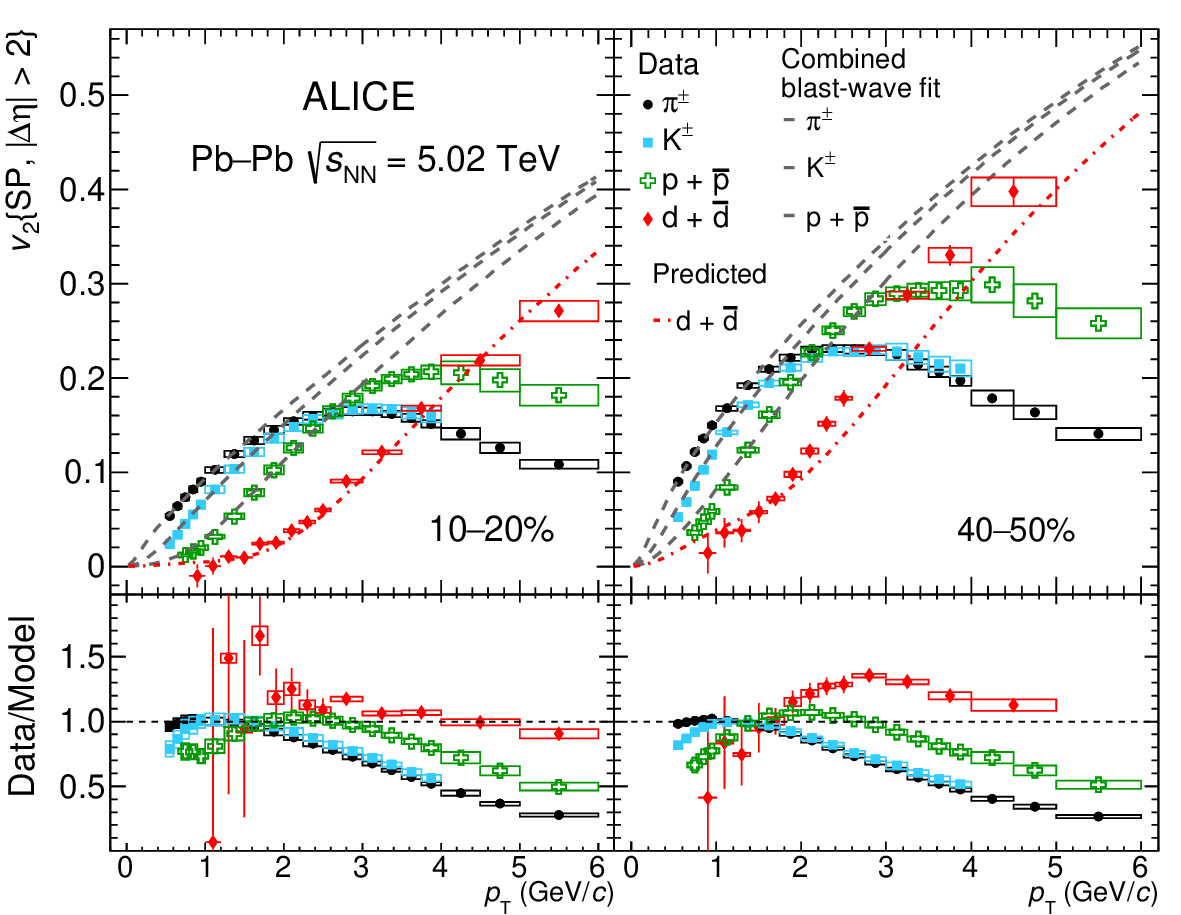}
\caption{(Color online) Blast-wave fits to the \vtwo(\pt) of pions, kaons,  and protons~\cite{identifiedHadronsFlow} and predictions of the deuterons \vtwo(\pt) for the centrality intervals 10--20$\%$ (left) and 40--50$\%$ (right). In the lower panels, the data-to-fit ratios are shown for pions, kaons,  and protons as well as the ratio of the deuterons \vtwo to the \mbox{blast-wave} predictions. Vertical bars and boxes sent the statistical and systematic uncertainties, respectively. The dashed line at one is to guide the eye.}
\label{fig:BWComparison}
\end{center}
\end{figure}

The blast-wave fits to the \vtwo\ of pions, kaons, and protons and the predictions for the deuterons \vtwo\ are reported in Fig.~\ref{fig:BWComparison} for the centrality intervals 10--20$\%$ and 40--50$\%$. In the lower panels, the data-to-fit ratios for pions, kaons, and protons and the ratios of the deuterons \vtwo\ to the model are shown. Because of the finite size of the \pt\ intervals, the average of the blast-wave function within the interval, weighted by the \pt~spectrum of the corresponding particle species, is considered in the calculation of these ratios.

\begin{figure}[!htb]
\begin{center}
\includegraphics[width=1\textwidth]{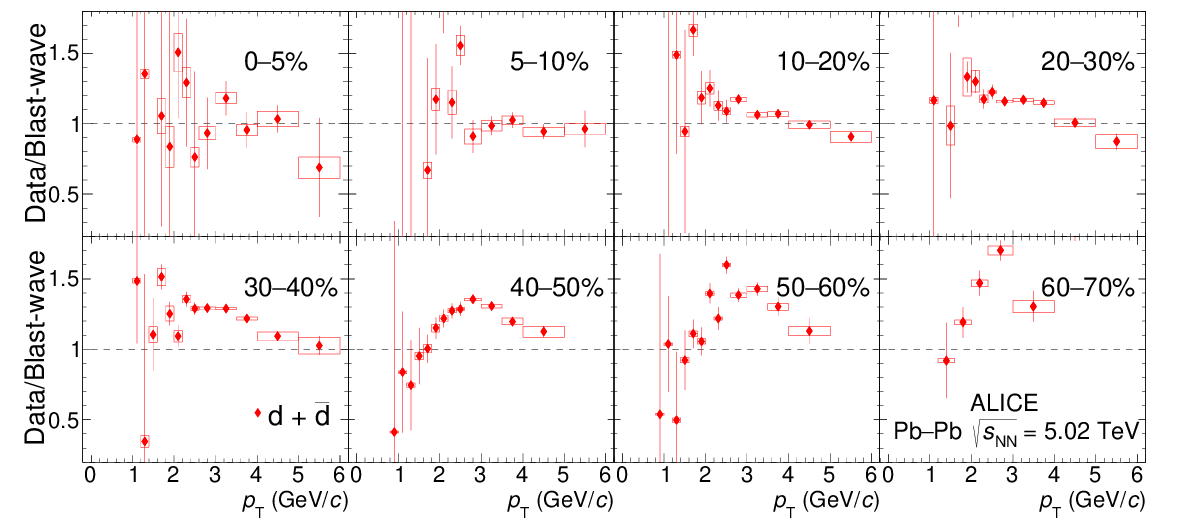}
\caption{(Color online) Data-to-model ratios of the deuterons \vtwo to the blast-wave predictions as a function of \pt\ for different centrality intervals as indicated in each pad. Vertical bars and boxes represent the statistical and systematic uncertainties, respectively.}
\label{fig:BWComparisonRatios}
\end{center}
\end{figure}

The predictions of the blast-wave model underestimate the deuterons elliptic flow experimental values in semi-peripheral collisions for \pt~$>$~1.4~\gmom, while they are close to the measurements for central events in the measured \pt\ interval. This is better observed in Fig.~\ref{fig:BWComparisonRatios}, which shows the centrality evolution of the data-to-model ratios. 

\subsection{Test of the coalescence hypothesis}
\label{subsec:CoalescenceComparison}
The deuterons \vtwo\ and \vthree\ are compared to the expectations of a coalescence approach based on mass number scaling and isospin symmetry, for which the proton and neutron $v_{2}$ ($v_{3}$) are identical.  
In particular, the  \vtwo(\vthree) measured for protons~\cite{identifiedHadronsFlow} was used to predict the \vtwo(\vthree) of deuterons using the following relation~\cite{Molnar:2003ff}
\begin{equation}
v_{2(3),\mathrm{d}}(p_{\rm T}) = \frac{2 v_{2(3),\mathrm{p}} (p_{\rm {T}}/2)}{1+2{v^2_{2(3),\rm{p}}} (p_{\mathrm {T}}/2)}.
\label{eq:v2coal}
\end{equation} 

\begin{figure}[!htb]
\begin{center}
\includegraphics[width=0.49\textwidth]{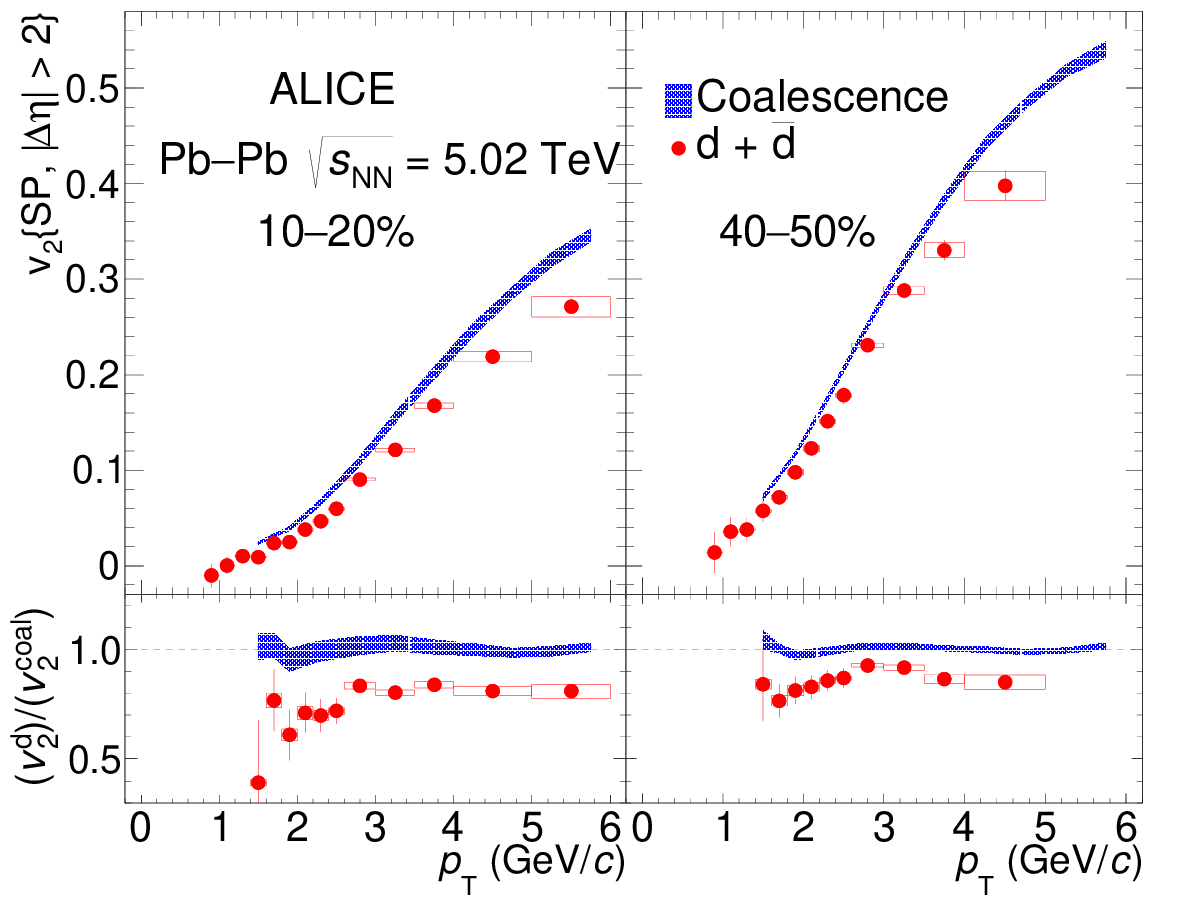}
\includegraphics[width=0.49\textwidth]{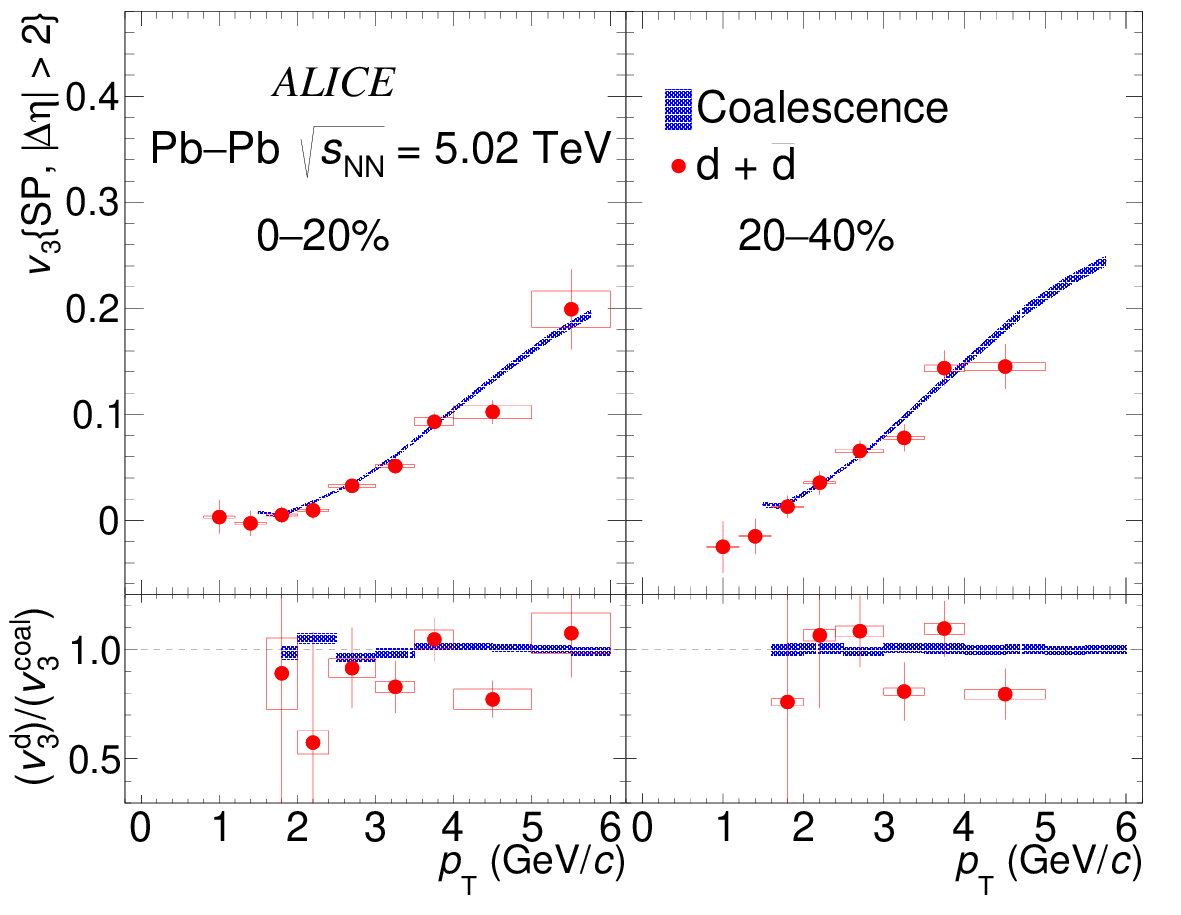}
\caption{(Color online) Measured deuterons \vtwo\ and \vthree\ (red circles) compared with the expectations from simple 
coalescence (Eq.~\ref{eq:v2coal}) (blue shaded bands) for two centrality intervals. In the left panel, the \vtwo\ measurements in the 10--20$\%$ and 40--50$\%$ centrality intervals is shown. The right panel displays the results of \vthree\ in the 0--20$\%$ and 20--40$\%$ centrality intervals. The bottom panels show the ratio between the measured \vtwo(\vthree) and the expectations from the coalescence model. In each panel, vertical bars and boxes represent the statistical and systematic uncertainties, respectively. The line at one is to guide the eye. 
}
\label{fig:CoalComparisonv2v3}
\end{center}
\end{figure}

The results of this calculation for different centrality intervals for \vtwo are shown in the left panel of Fig.~\ref{fig:CoalComparisonv2v3}. The measured elliptic flow in 10--20\% and 40--50\% centrality intervals of deuterons is compared with coalescence model predictions from Eq.~\ref{eq:v2coal} using the measured $v_{n}$ of protons. Similarly, the right panel of Fig.~\ref{fig:CoalComparisonv2v3} shows a comparison between the calculated and measured \vthree\ in the 0--20\% and 20--40\% centrality intervals.

\begin{figure}[!htb]
\begin{center}
\includegraphics[width=1\textwidth]{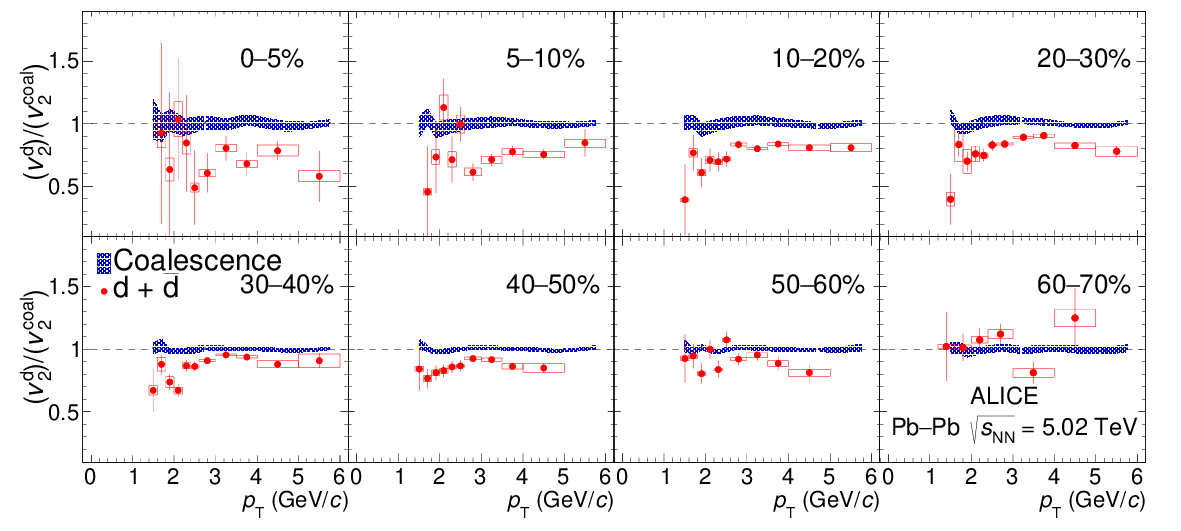}
\caption{(Color online) Centrality evolution of the deuterons \vtwo\ compared with the expectations from simple coalescence model~(Eq.~\ref{eq:v2coal}). 
Vertical bars and boxes represent the statistical and systematic uncertainties, respectively.}
\label{fig:CoalComparisonv2Ratios}
\end{center}
\end{figure}

The coalescence model overestimates the deuteron \vtwo\ by about 20$\%$ to 30$\%$ in central collisions and is close to the data for semi-peripheral collisions, as illustrated in Fig.~\ref{fig:CoalComparisonv2Ratios} which shows the centrality evolution of the data-to-model ratio. 
The coalescence approach seems to have a slightly better agreement with deuterons \vthree; however, the large statistical uncertainties on the \vthree\ measurements do not allow for conclusive statements.

\subsection{Comparison with iEBE-VISHNU and coalescence calculations}
\label{subsec:comparison_iEBEVISHNU}

\begin{figure}[!htb]
\begin{tabular}{c}
\includegraphics[width=0.45\textwidth]{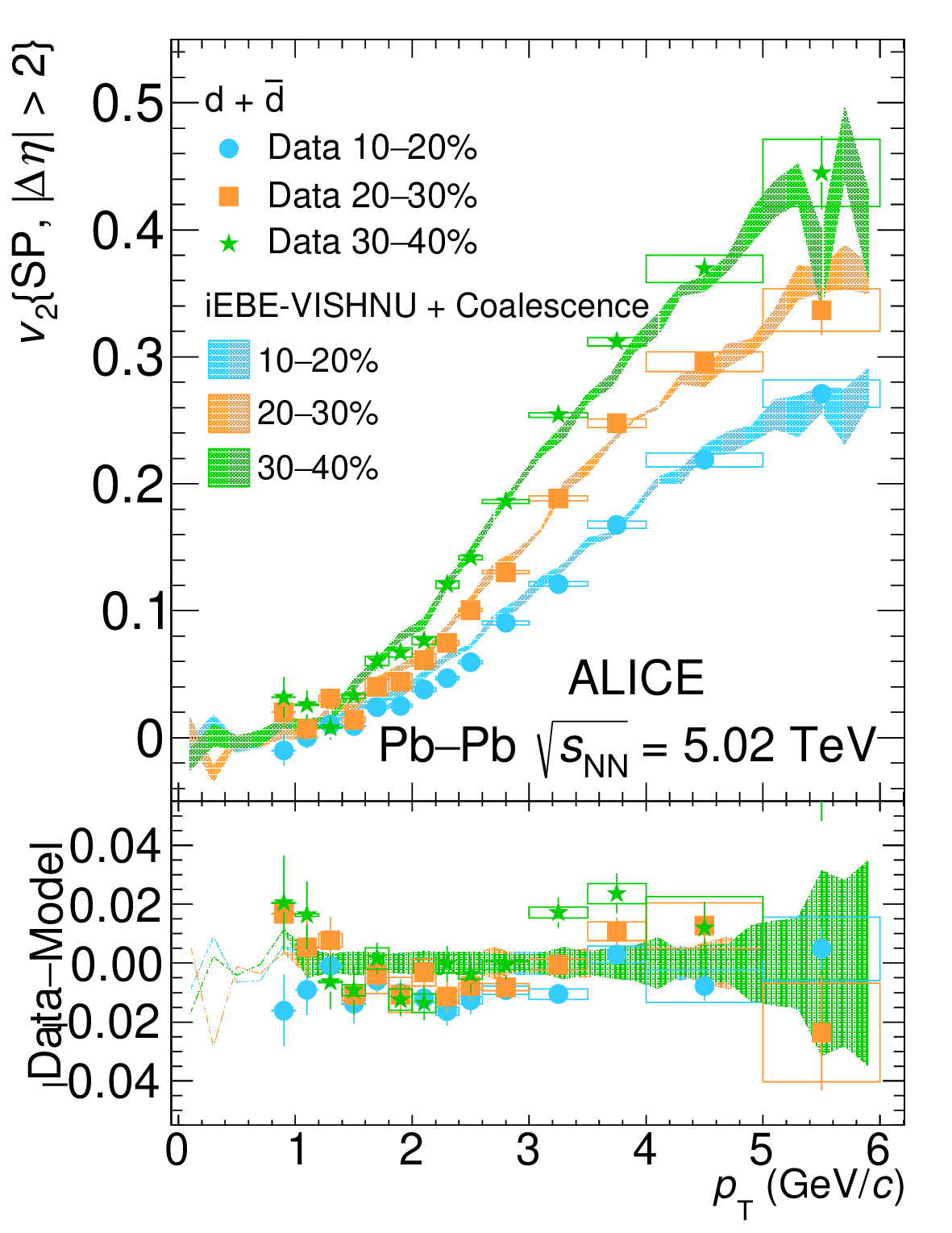}
\includegraphics[width=0.45\textwidth]{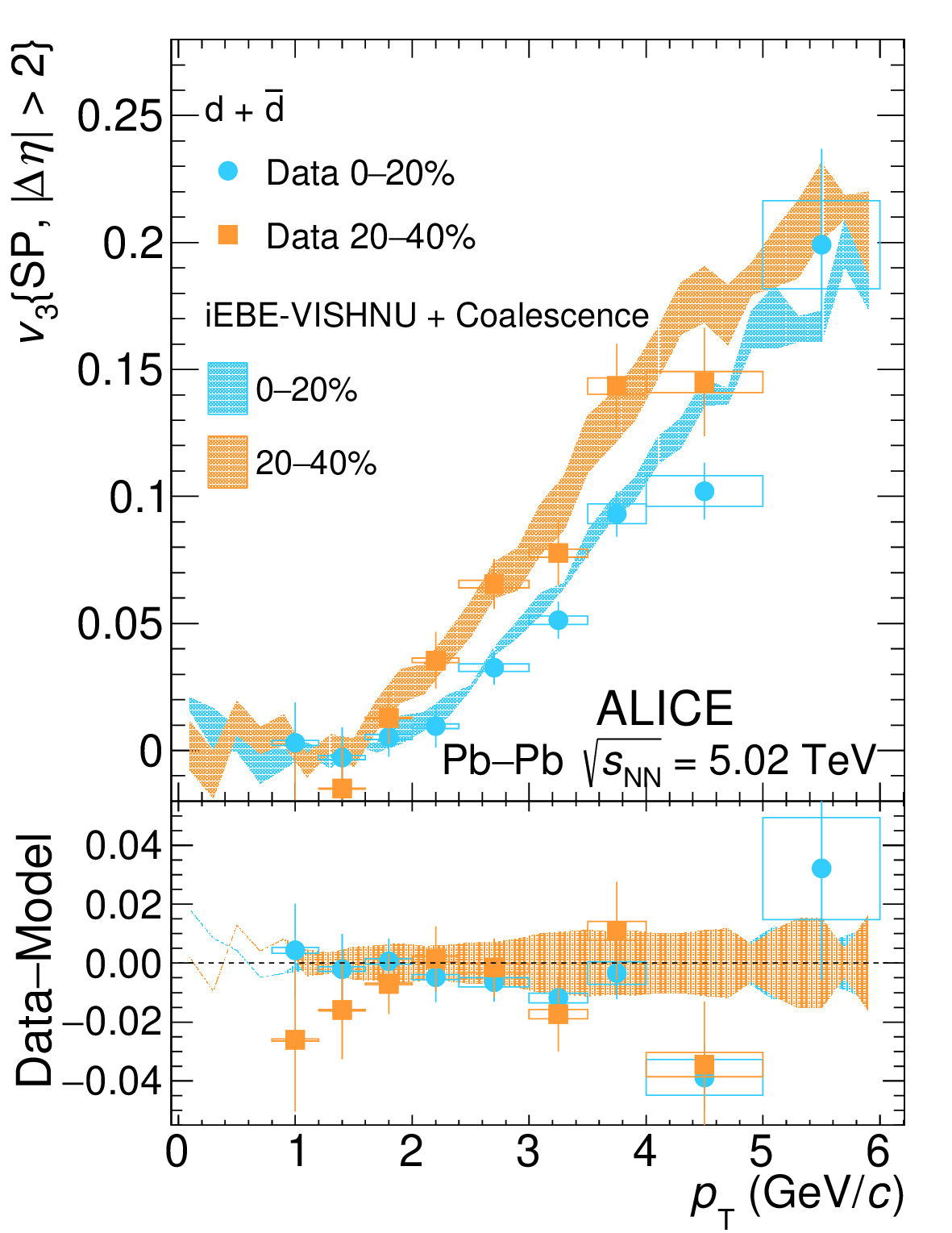}
\end{tabular}
\caption{Elliptic (left) and triangular (right) flow of deuterons compared to the predictions 
iEBE-VISHNU hybrid model with AMPT initial conditions~\cite{iEBE_VISHNU}. 
The predictions are shown as bands whose widths represent the statistical uncertainties associated with the model.
The data-to-model differences are shown in the lower panels. Vertical bars and boxes represent the statistical and systematic uncertainties, respectively. }
\label{fig:iEBEVISHNU}
\end{figure}

In  Fig.~\ref{fig:iEBEVISHNU}, the deuterons \vtwo\ and \vthree\ are compared to a model~\cite{iEBE_VISHNU} implementing light nuclei formation via coalescence of nucleons originating from a hydrodynamical evolution of the fireball coupled to an UrQMD simulation of the hadronic cascade~\mbox{\cite{UrQMD1, UrQMD2}}. 
In this model, the coalescence probability is calculated as the superposition of the wave functions of protons and neutrons and the Wigner function of the deuterons. The coalescence happens in a flowing medium introducing position-momentum correlations, which are  absent in the simple coalescence approach. The phase-space distributions of protons and neutrons are generated from the iEBE-VISHNU hybrid model with AMPT~\cite{Lin:2004en} initial conditions. 
This model provides a good description of the protons spectra up to 3~\gmom\ and of the deuterons \vtwo\ measured in \PbPb collisions at \snn~=~2.76~TeV~\cite{iEBE_VISHNU}.
The predictions are consistent with the measured deuterons \vtwo\ for events with centrality larger than 20\% and for measured \vthree\ within the statistical and systematical uncertainties, while some discrepancy at the level of 2$\sigma$ (taking into account statistical and systematical uncertainties in quadrature) are observed for for the centrality interval 10--20\% as shown in Fig.~\ref{fig:iEBEVISHNU}.

\subsection{Comparison with a hybrid (hydrodynamics + transport) approach expectations}
\label{subsec:comparison_microscopicApproach}

\begin{figure}[!htb]
\begin{center}
\includegraphics[width=0.5\textwidth]{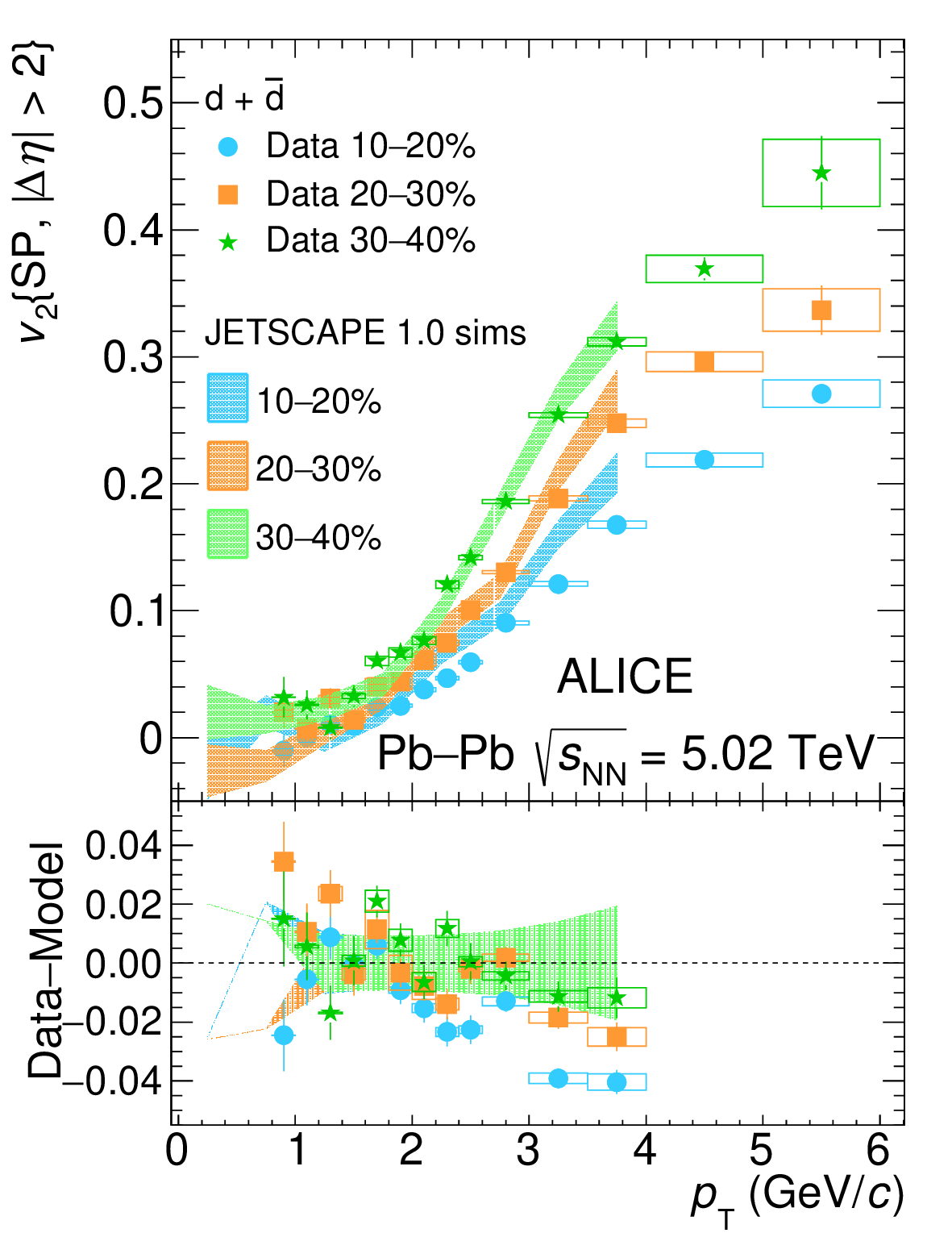}
\caption{Measured deuterons \vtwo\ compared to the predictions from a microscopic model~\cite{HydroSMASH} based on the JETSCAPE generator~\cite{JETSCAPE}. The model predictions, based on SMASH  afterburner and which used TRENTo~\cite{Trento} initial conditions, are shown as bands. The width of the band represents the statistical uncertainty associated with the model. In the lower panel the data-to-model differences are shown. Vertical bars and boxes represent the statistical and systematic uncertainties, respectively.}
\label{fig:SMASHcomparison}
\end{center}
\end{figure}

The deuterons \vtwo\ measured in the centrality intervals 10--20$\%$, 20--30$\%$, and 30--40$\%$ is compared in Fig.~\ref{fig:SMASHcomparison} with the predictions from a hybrid model based on relativistic viscous hydrodynamics, with fluctuating initial conditions generated by T$_\mathrm{R}$ENTo~\cite{Trento}, coupled to the hadronic afterburner SMASH~\cite{HydroSMASH}. The simulations are obtained by using the JETSCAPE 1.0 event generator~\cite{JETSCAPE}. 
The parameters of this model, including the shear and bulk viscosities, are tuned to the measurements of \pt spectra and azimuthal flow of pions, kaons, and protons obtained by ALICE in \PbPb collisions at \snn~=~2.76~TeV~\cite{SpectraIdentifiedHadrons276,FlowIdentifiedHadrons276} and by PHENIX and STAR in Au--Au collisions at \snn~=~200~GeV~\mbox{\cite{SpectraIdentifiedHadrons200GeVPHENIX,SpectraIdentifiedHadrons200GeVSTAR,FlowIdentifiedHadrons200GeV}}. 
The interactions of deuterons with other hadrons in the hadron gas phase are simulated using SMASH in which all known resonances and the experimentally known cross sections, most importantly $\pi$d $\rightarrow \pi$np and its inverse reaction, are included.

In this model, the number of deuterons at the kinetic freezeout is independent from their primordial abundance at the Cooper-Frye hypersurface. It was found that even when their initial number is set to zero, the number of deuterons regenerated in the hadronic phase converges towards the equilibrium value, which is the same as that predicted by the statistical hadronization model. 
Considering that in this model only $\sim 1\%$ of the primordial deuterons survive the hadronic stage, the elliptic flow of deuterons observed after the kinetic freezeout is almost identical to that of the regenerated ones. 
For this reason, deuterons are not sampled at the Cooper-Frye hypersurface for these predictions. 

The model predictions are consistent with the measured \vtwo\ within the uncertainties in the centrality intervals 20--30\% and 30--40\% for 0.8$<$~\pt~$<$~4~\gmom, while the data are overestimated by up to 30\% in the centrality interval 10--20\% for ~\pt~$>$~2~\gmom.

\section{Summary}
\label{sec:conclusions}

The measurements of the \mbox{deuterons} \vtwo\ and the first measurement of  \vthree\ in \PbPb collisions at \snn = 5.02~TeV are presented. The observed centrality and \pt dependence are consistent with the expectations from relativistic hydrodynamics. A mass ordering is observed for \pt~$<$~3~\gmom\ when comparing these results with the measured \vtwo\ and \vthree\ of pions, kaons, and protons. The shift of the deuterons \vtwo towards higher \pt with respect to the measurement in \PbPb collisions at \snn~=~2.76~TeV, mainly due to a stronger radial flow at higher center-of-mass energy, is consistent with that observed for the proton \vtwo measurement.

The results of this measurement are compared with the expectations from the  simple coalescence approach, in which the deuterons \vtwo is obtained from that of protons assuming that the deuterons invariant yield is proportional to that of protons squared, and with the predictions of the blast-wave model. The deuterons \vtwo is overestimated by a simple coalescence approach, which describes the data only in peripheral (centrality $>$~50\%) collisions. On the other hand, the blast-wave model underestimates the peripheral measurements and it is close to the data in central collisions. These results are consistent with the scenario previously seen for deuterons and $^{3}$He elliptic flow: these simplified models bracket a region where the light nuclei \vtwo is 
located and describe reasonably the data in different multiplicity regimes, indicating that none of these two models is able to describe the deuterons production measurement from low to high multiplicity environments. 

Similar considerations are valid for the deuterons \vthree with some limitations due to the rather large statistical uncertainties. This specific aspect will be addressed with the larger data sample that will be collected in Run 3 following the ALICE upgrade, where a significant improvement of the statistical precision is expected. This measurement will be crucial to better constrain models that describe the production of light nuclei in heavy-ion collisions.  

A more advanced coalescence model coupled to hydrodynamics and the hadronic afterburner UrQMD, which takes into account the quantum-mechanical properties of nucleons and nuclei and space-momentum correlations of nucleons, provides a good description of the deuterons \vtwo and \vthree for \pt~$>$~2.5~\gmom. 
The model predictions deviate from the data at lower \pt, in particular for the centrality interval 10--20\%. The same model provides a good description of the deuterons \vtwo measured in \PbPb collisions at \snn~=~2.76~TeV and that of $^{3}$He at \snn~=~5.02~TeV.  
The deuterons \vtwo\ is also compared to the predictions from a hybrid model based on relativistic hydrodynamics coupled to the hadronic afterburner SMASH. The model predictions are consistent with the data within the uncertainties in the centrality intervals 20--30\% and 30--40\%, while a deviation of up to 30\% is observed in the centrality interval 10--20\% for 2~$<$~\pt~$<$~3~\gmom. 

In general, the state-of-the-art implementations of coalescence and the hybrid approach based on hydrodynamics coupled to hadronic afterburners 
provide better descriptions of the data compared to the simple coalescence and blast-wave models. 
Further efforts, both on experimental and theoretical side, are needed to have a more comprehensive 
understanding of dynamics and production of light nuclei.


\newenvironment{acknowledgement}{\relax}{\relax}
\begin{acknowledgement}
\section*{Acknowledgements}


The ALICE Collaboration would like to thank all its engineers and technicians for their invaluable contributions to the construction of the experiment and the CERN accelerator teams for the outstanding performance of the LHC complex.
The ALICE Collaboration gratefully acknowledges the resources and support provided by all Grid centres and the Worldwide LHC Computing Grid (WLCG) collaboration.
The ALICE Collaboration acknowledges the following funding agencies for their support in building and running the ALICE detector:
A. I. Alikhanyan National Science Laboratory (Yerevan Physics Institute) Foundation (ANSL), State Committee of Science and World Federation of Scientists (WFS), Armenia;
Austrian Academy of Sciences, Austrian Science Fund (FWF): [M 2467-N36] and Nationalstiftung f\"{u}r Forschung, Technologie und Entwicklung, Austria;
Ministry of Communications and High Technologies, National Nuclear Research Center, Azerbaijan;
Conselho Nacional de Desenvolvimento Cient\'{\i}fico e Tecnol\'{o}gico (CNPq), Financiadora de Estudos e Projetos (Finep), Funda\c{c}\~{a}o de Amparo \`{a} Pesquisa do Estado de S\~{a}o Paulo (FAPESP) and Universidade Federal do Rio Grande do Sul (UFRGS), Brazil;
Ministry of Education of China (MOEC) , Ministry of Science \& Technology of China (MSTC) and National Natural Science Foundation of China (NSFC), China;
Ministry of Science and Education and Croatian Science Foundation, Croatia;
Centro de Aplicaciones Tecnol\'{o}gicas y Desarrollo Nuclear (CEADEN), Cubaenerg\'{\i}a, Cuba;
Ministry of Education, Youth and Sports of the Czech Republic, Czech Republic;
The Danish Council for Independent Research | Natural Sciences, the VILLUM FONDEN and Danish National Research Foundation (DNRF), Denmark;
Helsinki Institute of Physics (HIP), Finland;
Commissariat \`{a} l'Energie Atomique (CEA) and Institut National de Physique Nucl\'{e}aire et de Physique des Particules (IN2P3) and Centre National de la Recherche Scientifique (CNRS), France;
Bundesministerium f\"{u}r Bildung und Forschung (BMBF) and GSI Helmholtzzentrum f\"{u}r Schwerionenforschung GmbH, Germany;
General Secretariat for Research and Technology, Ministry of Education, Research and Religions, Greece;
National Research, Development and Innovation Office, Hungary;
Department of Atomic Energy Government of India (DAE), Department of Science and Technology, Government of India (DST), University Grants Commission, Government of India (UGC) and Council of Scientific and Industrial Research (CSIR), India;
Indonesian Institute of Science, Indonesia;
Centro Fermi - Museo Storico della Fisica e Centro Studi e Ricerche Enrico Fermi and Istituto Nazionale di Fisica Nucleare (INFN), Italy;
Institute for Innovative Science and Technology , Nagasaki Institute of Applied Science (IIST), Japanese Ministry of Education, Culture, Sports, Science and Technology (MEXT) and Japan Society for the Promotion of Science (JSPS) KAKENHI, Japan;
Consejo Nacional de Ciencia (CONACYT) y Tecnolog\'{i}a, through Fondo de Cooperaci\'{o}n Internacional en Ciencia y Tecnolog\'{i}a (FONCICYT) and Direcci\'{o}n General de Asuntos del Personal Academico (DGAPA), Mexico;
Nederlandse Organisatie voor Wetenschappelijk Onderzoek (NWO), Netherlands;
The Research Council of Norway, Norway;
Commission on Science and Technology for Sustainable Development in the South (COMSATS), Pakistan;
Pontificia Universidad Cat\'{o}lica del Per\'{u}, Peru;
Ministry of Science and Higher Education, National Science Centre and WUT ID-UB, Poland;
Korea Institute of Science and Technology Information and National Research Foundation of Korea (NRF), Republic of Korea;
Ministry of Education and Scientific Research, Institute of Atomic Physics and Ministry of Research and Innovation and Institute of Atomic Physics, Romania;
Joint Institute for Nuclear Research (JINR), Ministry of Education and Science of the Russian Federation, National Research Centre Kurchatov Institute, Russian Science Foundation and Russian Foundation for Basic Research, Russia;
Ministry of Education, Science, Research and Sport of the Slovak Republic, Slovakia;
National Research Foundation of South Africa, South Africa;
Swedish Research Council (VR) and Knut \& Alice Wallenberg Foundation (KAW), Sweden;
European Organization for Nuclear Research, Switzerland;
Suranaree University of Technology (SUT), National Science and Technology Development Agency (NSDTA) and Office of the Higher Education Commission under NRU project of Thailand, Thailand;
Turkish Atomic Energy Agency (TAEK), Turkey;
National Academy of  Sciences of Ukraine, Ukraine;
Science and Technology Facilities Council (STFC), United Kingdom;
National Science Foundation of the United States of America (NSF) and United States Department of Energy, Office of Nuclear Physics (DOE NP), United States of America.
\end{acknowledgement}


\bibliographystyle{utphys}   
\bibliography{bibliography}

\newpage
\appendix

%
%

\section{The ALICE Collaboration}
\label{app:collab}

\begingroup
\small
\begin{flushleft}
S.~Acharya\Irefn{org141}\And 
D.~Adamov\'{a}\Irefn{org95}\And 
A.~Adler\Irefn{org74}\And 
J.~Adolfsson\Irefn{org81}\And 
M.M.~Aggarwal\Irefn{org100}\And 
G.~Aglieri Rinella\Irefn{org34}\And 
M.~Agnello\Irefn{org30}\And 
N.~Agrawal\Irefn{org10}\textsuperscript{,}\Irefn{org54}\And 
Z.~Ahammed\Irefn{org141}\And 
S.~Ahmad\Irefn{org16}
% ...
\And
\renewcommand\labelenumi{\textsuperscript{\theenumi}~}

\section*{Affiliation notes}
\renewcommand\theenumi{\roman{enumi}}
\begin{Authlist}
\item \Adef{org*}Deceased
\item \Adef{orgI}Italian National Agency for New Technologies, Energy and Sustainable Economic Development (ENEA), Bologna, Italy
\item \Adef{orgII}Dipartimento DET del Politecnico di Torino, Turin, Italy
\item \Adef{orgIII}M.V. Lomonosov Moscow State University, D.V. Skobeltsyn Institute of Nuclear, Physics, Moscow, Russia
\item \Adef{orgIV}Department of Applied Physics, Aligarh Muslim University, Aligarh, India
\item \Adef{orgV}Institute of Theoretical Physics, University of Wroclaw, Poland
\end{Authlist}

\section*{Collaboration Institutes}
\renewcommand\theenumi{\arabic{enumi}~}
\begin{Authlist}
\item \Idef{org1}A.I. Alikhanyan National Science Laboratory (Yerevan Physics Institute) Foundation, Yerevan, Armenia
\item \Idef{org2}Bogolyubov Institute for Theoretical Physics, National Academy of Sciences of Ukraine, Kiev, Ukraine
\item \Idef{org3}Bose Institute, Department of Physics  and Centre for Astroparticle Physics and Space Science (CAPSS), Kolkata, India
\item \Idef{org4}Budker Institute for Nuclear Physics, Novosibirsk, Russia
\item \Idef{org5}California Polytechnic State University, San Luis Obispo, California, United States
\item \Idef{org6}Central China Normal University, Wuhan, China
\item \Idef{org7}Centre de Calcul de l'IN2P3, Villeurbanne, Lyon, France
\item \Idef{org8}Centro de Aplicaciones Tecnol\'{o}gicas y Desarrollo Nuclear (CEADEN), Havana, Cuba
\item \Idef{org9}Centro de Investigaci\'{o}n y de Estudios Avanzados (CINVESTAV), Mexico City and M\'{e}rida, Mexico
\item \Idef{org10}Centro Fermi - Museo Storico della Fisica e Centro Studi e Ricerche ``Enrico Fermi', Rome, Italy
\item \Idef{org11}Chicago State University, Chicago, Illinois, United States
\item \Idef{org12}China Institute of Atomic Energy, Beijing, China
\item \Idef{org13}Comenius University Bratislava, Faculty of Mathematics, Physics and Informatics, Bratislava, Slovakia
\item \Idef{org14}COMSATS University Islamabad, Islamabad, Pakistan
\item \Idef{org15}Creighton University, Omaha, Nebraska, United States
\item \Idef{org16}Department of Physics, Aligarh Muslim University, Aligarh, India
\item \Idef{org17}Department of Physics, Pusan National University, Pusan, Republic of Korea
\item \Idef{org18}Department of Physics, Sejong University, Seoul, Republic of Korea
\item \Idef{org19}Department of Physics, University of California, Berkeley, California, United States
\item \Idef{org20}Department of Physics, University of Oslo, Oslo, Norway
\item \Idef{org21}Department of Physics and Technology, University of Bergen, Bergen, Norway
\item \Idef{org22}Dipartimento di Fisica dell'Universit\`{a} 'La Sapienza' and Sezione INFN, Rome, Italy
\item \Idef{org23}Dipartimento di Fisica dell'Universit\`{a} and Sezione INFN, Cagliari, Italy
\item \Idef{org24}Dipartimento di Fisica dell'Universit\`{a} and Sezione INFN, Trieste, Italy
\item \Idef{org25}Dipartimento di Fisica dell'Universit\`{a} and Sezione INFN, Turin, Italy
\item \Idef{org26}Dipartimento di Fisica e Astronomia dell'Universit\`{a} and Sezione INFN, Bologna, Italy
\item \Idef{org27}Dipartimento di Fisica e Astronomia dell'Universit\`{a} and Sezione INFN, Catania, Italy
\item \Idef{org28}Dipartimento di Fisica e Astronomia dell'Universit\`{a} and Sezione INFN, Padova, Italy
\item \Idef{org29}Dipartimento di Fisica `E.R.~Caianiello' dell'Universit\`{a} and Gruppo Collegato INFN, Salerno, Italy
\item \Idef{org30}Dipartimento DISAT del Politecnico and Sezione INFN, Turin, Italy
\item \Idef{org31}Dipartimento di Scienze e Innovazione Tecnologica dell'Universit\`{a} del Piemonte Orientale and INFN Sezione di Torino, Alessandria, Italy
\item \Idef{org32}Dipartimento di Scienze MIFT, Universit\`{a} di Messina, Messina, Italy
\item \Idef{org33}Dipartimento Interateneo di Fisica `M.~Merlin' and Sezione INFN, Bari, Italy
\item \Idef{org34}European Organization for Nuclear Research (CERN), Geneva, Switzerland
\item \Idef{org35}Faculty of Electrical Engineering, Mechanical Engineering and Naval Architecture, University of Split, Split, Croatia
\item \Idef{org36}Faculty of Engineering and Science, Western Norway University of Applied Sciences, Bergen, Norway
\item \Idef{org37}Faculty of Nuclear Sciences and Physical Engineering, Czech Technical University in Prague, Prague, Czech Republic
\item \Idef{org38}Faculty of Science, P.J.~\v{S}af\'{a}rik University, Ko\v{s}ice, Slovakia
\item \Idef{org39}Frankfurt Institute for Advanced Studies, Johann Wolfgang Goethe-Universit\"{a}t Frankfurt, Frankfurt, Germany
\item \Idef{org40}Fudan University, Shanghai, China
\item \Idef{org41}Gangneung-Wonju National University, Gangneung, Republic of Korea
\item \Idef{org42}Gauhati University, Department of Physics, Guwahati, India
\item \Idef{org43}Helmholtz-Institut f\"{u}r Strahlen- und Kernphysik, Rheinische Friedrich-Wilhelms-Universit\"{a}t Bonn, Bonn, Germany
\item \Idef{org44}Helsinki Institute of Physics (HIP), Helsinki, Finland
\item \Idef{org45}High Energy Physics Group,  Universidad Aut\'{o}noma de Puebla, Puebla, Mexico
\item \Idef{org46}Hiroshima University, Hiroshima, Japan
\item \Idef{org47}Hochschule Worms, Zentrum  f\"{u}r Technologietransfer und Telekommunikation (ZTT), Worms, Germany
\item \Idef{org48}Horia Hulubei National Institute of Physics and Nuclear Engineering, Bucharest, Romania
\item \Idef{org49}Indian Institute of Technology Bombay (IIT), Mumbai, India
\item \Idef{org50}Indian Institute of Technology Indore, Indore, India
\item \Idef{org51}Indonesian Institute of Sciences, Jakarta, Indonesia
\item \Idef{org52}INFN, Laboratori Nazionali di Frascati, Frascati, Italy
\item \Idef{org53}INFN, Sezione di Bari, Bari, Italy
\item \Idef{org54}INFN, Sezione di Bologna, Bologna, Italy
\item \Idef{org55}INFN, Sezione di Cagliari, Cagliari, Italy
\item \Idef{org56}INFN, Sezione di Catania, Catania, Italy
\item \Idef{org57}INFN, Sezione di Padova, Padova, Italy
\item \Idef{org58}INFN, Sezione di Roma, Rome, Italy
\item \Idef{org59}INFN, Sezione di Torino, Turin, Italy
\item \Idef{org60}INFN, Sezione di Trieste, Trieste, Italy
\item \Idef{org61}Inha University, Incheon, Republic of Korea
\item \Idef{org62}Institute for Nuclear Research, Academy of Sciences, Moscow, Russia
\item \Idef{org63}Institute for Subatomic Physics, Utrecht University/Nikhef, Utrecht, Netherlands
\item \Idef{org64}Institute of Experimental Physics, Slovak Academy of Sciences, Ko\v{s}ice, Slovakia
\item \Idef{org65}Institute of Physics, Homi Bhabha National Institute, Bhubaneswar, India
\item \Idef{org66}Institute of Physics of the Czech Academy of Sciences, Prague, Czech Republic
\item \Idef{org67}Institute of Space Science (ISS), Bucharest, Romania
\item \Idef{org68}Institut f\"{u}r Kernphysik, Johann Wolfgang Goethe-Universit\"{a}t Frankfurt, Frankfurt, Germany
\item \Idef{org69}Instituto de Ciencias Nucleares, Universidad Nacional Aut\'{o}noma de M\'{e}xico, Mexico City, Mexico
\item \Idef{org70}Instituto de F\'{i}sica, Universidade Federal do Rio Grande do Sul (UFRGS), Porto Alegre, Brazil
\item \Idef{org71}Instituto de F\'{\i}sica, Universidad Nacional Aut\'{o}noma de M\'{e}xico, Mexico City, Mexico
\item \Idef{org72}iThemba LABS, National Research Foundation, Somerset West, South Africa
\item \Idef{org73}Jeonbuk National University, Jeonju, Republic of Korea
\item \Idef{org74}Johann-Wolfgang-Goethe Universit\"{a}t Frankfurt Institut f\"{u}r Informatik, Fachbereich Informatik und Mathematik, Frankfurt, Germany
\item \Idef{org75}Joint Institute for Nuclear Research (JINR), Dubna, Russia
\item \Idef{org76}Korea Institute of Science and Technology Information, Daejeon, Republic of Korea
\item \Idef{org77}KTO Karatay University, Konya, Turkey
\item \Idef{org78}Laboratoire de Physique des 2 Infinis, Ir\`{e}ne Joliot-Curie, Orsay, France
\item \Idef{org79}Laboratoire de Physique Subatomique et de Cosmologie, Universit\'{e} Grenoble-Alpes, CNRS-IN2P3, Grenoble, France
\item \Idef{org80}Lawrence Berkeley National Laboratory, Berkeley, California, United States
\item \Idef{org81}Lund University Department of Physics, Division of Particle Physics, Lund, Sweden
\item \Idef{org82}Nagasaki Institute of Applied Science, Nagasaki, Japan
\item \Idef{org83}Nara Women{'}s University (NWU), Nara, Japan
\item \Idef{org84}National and Kapodistrian University of Athens, School of Science, Department of Physics , Athens, Greece
\item \Idef{org85}National Centre for Nuclear Research, Warsaw, Poland
\item \Idef{org86}National Institute of Science Education and Research, Homi Bhabha National Institute, Jatni, India
\item \Idef{org87}National Nuclear Research Center, Baku, Azerbaijan
\item \Idef{org88}National Research Centre Kurchatov Institute, Moscow, Russia
\item \Idef{org89}Niels Bohr Institute, University of Copenhagen, Copenhagen, Denmark
\item \Idef{org90}Nikhef, National institute for subatomic physics, Amsterdam, Netherlands
\item \Idef{org91}NRC Kurchatov Institute IHEP, Protvino, Russia
\item \Idef{org92}NRC \guillemotleft Kurchatov\guillemotright~Institute - ITEP, Moscow, Russia
\item \Idef{org93}NRNU Moscow Engineering Physics Institute, Moscow, Russia
\item \Idef{org94}Nuclear Physics Group, STFC Daresbury Laboratory, Daresbury, United Kingdom
\item \Idef{org95}Nuclear Physics Institute of the Czech Academy of Sciences, \v{R}e\v{z} u Prahy, Czech Republic
\item \Idef{org96}Oak Ridge National Laboratory, Oak Ridge, Tennessee, United States
\item \Idef{org97}Ohio State University, Columbus, Ohio, United States
\item \Idef{org98}Petersburg Nuclear Physics Institute, Gatchina, Russia
\item \Idef{org99}Physics department, Faculty of science, University of Zagreb, Zagreb, Croatia
\item \Idef{org100}Physics Department, Panjab University, Chandigarh, India
\item \Idef{org101}Physics Department, University of Jammu, Jammu, India
\item \Idef{org102}Physics Department, University of Rajasthan, Jaipur, India
\item \Idef{org103}Physikalisches Institut, Eberhard-Karls-Universit\"{a}t T\"{u}bingen, T\"{u}bingen, Germany
\item \Idef{org104}Physikalisches Institut, Ruprecht-Karls-Universit\"{a}t Heidelberg, Heidelberg, Germany
\item \Idef{org105}Physik Department, Technische Universit\"{a}t M\"{u}nchen, Munich, Germany
\item \Idef{org106}Politecnico di Bari, Bari, Italy
\item \Idef{org107}Research Division and ExtreMe Matter Institute EMMI, GSI Helmholtzzentrum f\"ur Schwerionenforschung GmbH, Darmstadt, Germany
\item \Idef{org108}Rudjer Bo\v{s}kovi\'{c} Institute, Zagreb, Croatia
\item \Idef{org109}Russian Federal Nuclear Center (VNIIEF), Sarov, Russia
\item \Idef{org110}Saha Institute of Nuclear Physics, Homi Bhabha National Institute, Kolkata, India
\item \Idef{org111}School of Physics and Astronomy, University of Birmingham, Birmingham, United Kingdom
\item \Idef{org112}Secci\'{o}n F\'{\i}sica, Departamento de Ciencias, Pontificia Universidad Cat\'{o}lica del Per\'{u}, Lima, Peru
\item \Idef{org113}St. Petersburg State University, St. Petersburg, Russia
\item \Idef{org114}Stefan Meyer Institut f\"{u}r Subatomare Physik (SMI), Vienna, Austria
\item \Idef{org115}SUBATECH, IMT Atlantique, Universit\'{e} de Nantes, CNRS-IN2P3, Nantes, France
\item \Idef{org116}Suranaree University of Technology, Nakhon Ratchasima, Thailand
\item \Idef{org117}Technical University of Ko\v{s}ice, Ko\v{s}ice, Slovakia
\item \Idef{org118}The Henryk Niewodniczanski Institute of Nuclear Physics, Polish Academy of Sciences, Cracow, Poland
\item \Idef{org119}The University of Texas at Austin, Austin, Texas, United States
\item \Idef{org120}Universidad Aut\'{o}noma de Sinaloa, Culiac\'{a}n, Mexico
\item \Idef{org121}Universidade de S\~{a}o Paulo (USP), S\~{a}o Paulo, Brazil
\item \Idef{org122}Universidade Estadual de Campinas (UNICAMP), Campinas, Brazil
\item \Idef{org123}Universidade Federal do ABC, Santo Andre, Brazil
\item \Idef{org124}University of Cape Town, Cape Town, South Africa
\item \Idef{org125}University of Houston, Houston, Texas, United States
\item \Idef{org126}University of Jyv\"{a}skyl\"{a}, Jyv\"{a}skyl\"{a}, Finland
\item \Idef{org127}University of Liverpool, Liverpool, United Kingdom
\item \Idef{org128}University of Science and Technology of China, Hefei, China
\item \Idef{org129}University of South-Eastern Norway, Tonsberg, Norway
\item \Idef{org130}University of Tennessee, Knoxville, Tennessee, United States
\item \Idef{org131}University of the Witwatersrand, Johannesburg, South Africa
\item \Idef{org132}University of Tokyo, Tokyo, Japan
\item \Idef{org133}University of Tsukuba, Tsukuba, Japan
\item \Idef{org134}Universit\'{e} Clermont Auvergne, CNRS/IN2P3, LPC, Clermont-Ferrand, France
\item \Idef{org135}Universit\'{e} de Lyon, Universit\'{e} Lyon 1, CNRS/IN2P3, IPN-Lyon, Villeurbanne, Lyon, France
\item \Idef{org136}Universit\'{e} de Strasbourg, CNRS, IPHC UMR 7178, F-67000 Strasbourg, France, Strasbourg, France
\item \Idef{org137}Universit\'{e} Paris-Saclay Centre d'Etudes de Saclay (CEA), IRFU, D\'{e}partment de Physique Nucl\'{e}aire (DPhN), Saclay, France
\item \Idef{org138}Universit\`{a} degli Studi di Foggia, Foggia, Italy
\item \Idef{org139}Universit\`{a} degli Studi di Pavia, Pavia, Italy
\item \Idef{org140}Universit\`{a} di Brescia, Brescia, Italy
\item \Idef{org141}Variable Energy Cyclotron Centre, Homi Bhabha National Institute, Kolkata, India
\item \Idef{org142}Warsaw University of Technology, Warsaw, Poland
\item \Idef{org143}Wayne State University, Detroit, Michigan, United States
\item \Idef{org144}Westf\"{a}lische Wilhelms-Universit\"{a}t M\"{u}nster, Institut f\"{u}r Kernphysik, M\"{u}nster, Germany
\item \Idef{org145}Wigner Research Centre for Physics, Budapest, Hungary
\item \Idef{org146}Yale University, New Haven, Connecticut, United States
\item \Idef{org147}Yonsei University, Seoul, Republic of Korea
\end{Authlist}
\endgroup
  
\end{document}